\documentclass[
reprint,
showpacs,
showkeys,
 amsmath,amssymb,
 aps,
 prc,
floatfix,
]{revtex4-1}

\usepackage[utf8]{inputenc}
\usepackage{graphicx}
\usepackage{color}
\usepackage{enumerate}
\usepackage[english]{babel}
\usepackage{paralist}
\usepackage{latexsym}
\usepackage{hyperref}

\usepackage{qmech-revtex}

\begin{document}

\preprint{}

\title{Dilepton production and reaction dynamics in heavy-ion collisions at SIS energies from coarse-grained transport simulations}

\author{Stephan Endres}
 \email{endres@th.physik.uni-frankfurt.de}
\author{Hendrik van Hees}%
\author{Janus Weil}%
\author{Marcus Bleicher}%
\affiliation{%
Frankfurt Institute for Advanced Studies,
Ruth-Moufang-Stra{\ss}e 1, D-60438 Frankfurt, Germany
}%
\affiliation{
Institut f{\"u}r Theoretische Physik, Universit{\"a}t Frankfurt,
Max-von-Laue-Stra{\ss}e 1, D-60438 Frankfurt, Germany
}

\date{July 6, 2015}

\begin{abstract}
  Dilepton invariant-mass spectra for heavy-ion collisions at SIS\,18
  and BEVALAC energies are calculated using a coarse-grained time
  evolution from the Ultra-relativistic Quantum Molecular Dynamics
  (UrQMD) model. The coarse-graining of the microscopic simulations
  enables to calculate thermal dilepton emission rates by application of
  in-medium spectral functions from equilibrium quantum-field
  theoretical calculations. The results show that extremely high baryon 
  chemical potentials dominate the evolution of the created hot 
  and dense fireball. Consequently, a significant modification of 
  the $\rho$ spectral shape becomes visible in the dilepton 
  invariant-mass spectrum, resulting in an enhancement in the
  low-mass region $M_{ee} = 200$ to 600\,MeV/$c^{2}$. This enhancement,
  mainly caused by baryonic effects on the $\rho$ spectral shape, can
  fully describe the experimentally observed excess above the hadronic
  cocktail contributions in Ar+KCl ($E_{\mathrm{lab}}=1.76$\,$A$GeV)
  reactions as measured by the HADES collaboration and also gives a good
  explanation of the older DLS Ca+Ca ($E_{\mathrm{lab}}=1.04$\,$A$GeV)
  data. For the larger Au+Au ($E_{\mathrm{lab}}=1.23$\,$A$GeV) system,
  we predict an even stronger excess from our calculations. A systematic
  comparison of the results for different system sizes from C+C to Au+Au
  shows that the thermal dilepton yield increases stronger
  ($\propto A^{4/3}$) than the hadronic background contributions, which
  scale with $A$, due to its sensitivity on the time evolution of the
  reaction. We stress that the findings of the present work are consistent 
  with our previous coarse-graining results for the NA60 measurements
  at top SPS energy. We argue that it is possible to describe the
  dilepton results from SIS\,18 up to SPS energies by considering the
  modifications of the $\rho$ spectral function inside a hot and dense
  medium within the same model.
\end{abstract}



\maketitle

\section{\label{intro} Introduction} 
The study of dilepton production has for long been proposed as a good
method to probe the change of hadronic properties in the hot and dense
matter created in heavy-ion collisions and also as a possible observable
for the creation of a deconfined phase at sufficiently high collision
energies \cite{Feinberg:1976ua,Shuryak:1978ij,Gale:1987ki,Gale:1987ey}. 
In contrast to the vacuum situation, a hadron can not only decay in a hot 
and dense medium but also interact with the constituents of the medium in
scattering processes and resonance excitation. Many theoretical efforts
have been undertaken over the last years to gain a better understanding
of the in-medium properties of vector mesons
\cite{Hatsuda:1991ez,Klingl:1997kf,Rapp:1999ej,Leupold:2009kz}. The
behavior of hadrons is of immanent interest for a full understanding of
the phase structure given by Quantum Chromodynamics (QCD). One
much-discussed aspect here is the change of the symmetries of QCD
which is expected when going from the vacuum to finite temperature and
finite baryochemical potential, especially the predicted restoration of chiral
symmetry \cite{Pisarski:1981mq,Brown:2001nh}. Unlike all hadronic
observables, which only provide information on the final freeze-out stage of
the system, dileptons are not subject to strong interactions and
consequently escape the fireball unscathed. However, this also means
that lepton pairs from all stages of the reaction will reach the
detector. Especially for a theoretical description this is a big
challenge, as it demands a realistic description of the whole space-time
evolution of the heavy-ion reaction and taking the various dilepton
sources into account.

On the experimental side several groups have undertaken the challenging
task to measure dilepton spectra in heavy-ion collisions and thereby
constrain the theoretical predictions. At SPS energies the NA60
Collaboration was able to measure the $\rho$ in-medium spectral function
for the first time, thanks to the high precision of the measurement
\cite{Arnaldi:2006jq}. The results were in line with previous CERES
results \cite{Agakishiev:1995xb} and found an excess in the invariant
mass range from 0.2 to 0.6\,GeV/$c^{2}$. This excess can be explained by
a broadening of the $\rho$ meson inside the hot and dense medium with
small mass shifts
\cite{vanHees:2006ng,vanHees:2007th,Dusling:2006yv,Endres:2014zua}. At
RHIC, these investigations were extended to even higher collision
energies with basically the same results except for less dominant
baryonic effects and a larger fraction of dileptons stemming from the QGP
\cite{Adamczyk:2013caa,STAR:2013tja}.

Still more challenging is the interpretation of the dilepton
measurements, which were performed in the low-energy regime at SIS\,18
and BEVALAC. For collision energies around
$E_{\mathrm{lab}} \approx 1-2$\,$A$GeV, which will be in the focus of
the present study, the DLS Collaboration measured a large excess beyond
the results of theoretical microscopic calculations several years
ago \cite{Porter:1997rc}. This disagreement between experiment and theory
was called the ``DLS puzzle''. In consequence, it triggered further
experimental and theoretical investigations. Recently, the HADES
experiment confirmed the former DLS results with a higher precision
\cite{Agakichiev:2006tg,Agakishiev:2007ts,
  Agakishiev:2009yf,Agakishiev:2011vf}. Although the theoretical
microscopic models have been largely improved and extended since this
time
\cite{Bratkovskaya:2007jk,Bratkovskaya:2008bf,Bratkovskaya:2013vx,Schmidt:2008hm,
  Weil:2012ji,Weil:2014rea}, a full and unambiguous description of the
data has not yet been found. A satisfying answer is complicated by the
fact that at such low energies a large number of processes contributes to the
dilepton production, for which many parameters (like cross-sections, 
branching ratios, etc.) are not well-known. In addition
interference effects are posing serious problems for transport
Monte-Carlo simulations based on the evolution of phase-space
densities. Here future measurements, for example in pion induced
reactions as conducted by HADES, could give better constraints for the
various parameters and reduce the uncertainty of the different
contributions \cite{Galatyuk:2014oaa}. In any case, a full description
of in-medium effects via off-shell dynamics or multi-particle
interactions at high densities remains a difficult task, although some
investigations on these issues have been conducted successfully
\cite{Schenke:2005ry,Schenke:2006uh,Schenke:2007zz,Bratkovskaya:1997mp,
  Bratkovskaya:2008bf,Cassing:2009vt,Barz:2009yz,Weil:2012qh}.

Besides the microscopic transport models, there exist also macroscopic
approaches describing the evolution of the heavy-ion collision in terms
of its thermodynamic properties. At high collision energies, i.e.~at SPS
or RHIC, a thermal calculation of dilepton emission is often applied,
where a fireball expansion or a hydrodynamic calculation is used to
model the bulk evolution of the system, while the dilepton emission is
calculated using spectral functions at a given $T$ and
$\mu_{\mathrm{B}}$
\cite{Dusling:2006yv,vanHees:2006ng,Ruppert:2007cr,Santini:2011zw,Vujanovic:2013jpa}.
But this approach works only if the collision energy is high
enough. The application at $E_\mathrm{lab} \approx 1$-2\,$A$GeV is
hardly reliable.

However, there are no reasons why a macroscopic description of the
reaction dynamics should not be possible at SIS\,18 and BEVALAC
energies, provided one can extract realistic values of energy and baryon
density and, in consequence, temperature and baryochemical potential. On
the contrary, due the the expected high values of the baryon chemical
potential, a study of the thermal properties of the system and the
influence on the spectral shape of vector mesons might be very
instructive. In the present work we argue that it is possible to obtain
realistic values of $T$ and $\mu_{\mathrm{B}}$ from microscopic calculations,
provided one uses a large ensemble of events and averages over them
(i.e.~one "coarse-grains" the results) to obtain a locally smooth
phase-space distribution. This ansatz, which was first presented and
applied for the calculation of dilepton and photon spectra in
\cite{Huovinen:2002im}, constitutes a compromise between
(non-equilibrium) microscopic transport simulations and the calculation
of dilepton emission with (equilibrium) spectral functions.

The same model was described in detail in \cite{Endres:2014zua} and
already successfully applied to investigate dilepton production at SPS
energies \cite{Endres:2014xga,Endres:2015wea}. The ansatz is used
basically unchanged for the present low-energy study. The only important
extension is the implementation of the in-medium spectral function for
the $\omega$ meson. While for the NA60 results the cocktail contribution of the $\omega$ was already subtracted from the thermal dilepton spectra, for comparing our calculations to the experimental HADES and DLS results a full description of the $\omega$ contribution is required. Furthermore, the high baryon densities and slow evolution of the system increase the significance of its medium modifications at the energies considered here.

Note that we focus the present investigation on the larger systems Ar+KCl and Au+Au, 
where local thermalization can be assumed due to the size of the
hot and dense fireball. For the smaller C+C system, which was also
studied by HADES and DLS (in nearly minimum bias reactions), one finds
that the average number of NN-collisions is so small that the assumption
of a local thermalization is questionable. Furthermore, the experimental
results indicate that the C+C dilepton spectra can be interpreted as the
superposition of the underlying p+p and p+n collisions without any
significant in-medium effects \cite{Agakishiev:2009yf}. (However, when
studying the effect of the system size on the dilepton production we
will also consider central C+C collisions later on for completeness.)

This paper is organized as follows. Section \ref{sec:1} gives an
introduction to the model and an overview of the different contributions
considered for the calculation. Section \ref{sec:2} then presents the
results for the thermodynamic evolution of the reaction (Sec.\
\ref{ssec:ReacDyn}) and the resulting dilepton spectra (Sec.\
\ref{ssec:DilSpec}), furthermore the effect of the system size and
fireball lifetime on the thermal dilepton yield is studied (Sec.\
\ref{syssize}). Finally, conclusions are drawn and an outlook to future
studies is given in section \ref{sec:3}.
\section{\label{sec:1} The Model} 
The full theoretical description of the dilepton spectra requires to
consider a large number of different production processes. At SIS
energies, all dileptons stem from hadronic sources, in contrast to the
situation at SPS, RHIC or LHC energies, where a significant contribution
is assumed to come from $q\bar{q}$-annihilation in the quark-gluon
plasma \cite{Rapp:2013nxa}. In general one can distinguish between two
different hadronic contributions, such from long-lived particles
(especially $\pi$ and $\eta$ mesons) and those from the short-lived
light vector mesons (mainly $\rho$, but also $\omega$ and $\phi$). The
former have a life time which is significantly larger than the duration
of the hot and dense stage in a heavy-ion collision. Consequently,
almost all of the decays of the long-lived mesons into lepton pairs 
will happen in the vacuum, i.e., no modification of the spectral 
shape is expected for those contributions. In contrast the latter 
mesons have a short life time and will therefore decay to a significant 
amount inside the fireball and their spectral properties 
are altered by the medium.

These differences are also significant for our approach. For the $\pi$
and $\eta$ meson contribution it is neither necessary nor adequate to
calculate thermal dilepton emission. The yields of these hadrons are
determined directly from calculations with the UrQMD transport
approach. For the $\rho$ and the $\omega$, however, we calculate the
thermal emission from coarse-grained transport simulations by application of
in-medium spectral functions. Here the influence of the space-time
evolution of the fireball is immanent, as the spectral shape will
largely depend on the values of temperature $T$ and baryon chemical
potential $\mu_{\mathrm{B}}$. Although for the $\phi$ some
medium-modifications of the spectral shape are predicted as well, we
here skip a full thermal calculation for the present investigation. On
the one hand, $\phi$ production is strongly suppressed at the low
energies considered here and will therefore hardly give any significant
contribution to the invariant mass spectrum, on the other hand the
predicted medium effects are rather small. Consequently, the $\phi$
contribution is directly extracted from the UrQMD calculations, as applied
for the $\pi$ and $\eta$. At higher invariant dilepton masses,
i.e.~mainly above 1 GeV/$c^{2}$, also multi-pion states in form of broad
resonances influence the dilepton production. This contribution is also
calculated as thermal emission.

The underlying model for all our considerations is the
Ultra-relativistic Quantum Molecular Dynamics approach (UrQMD), which is
a non-equilibrium cascade model
\cite{Bass:1998ca,Bleicher:1999xi,Petersen:2008dd,UrQMDweb}. It includes
all relevant hadronic resonances up to a mass of 2.2\,GeV/$c^{2}$. The
model gives an effective solution to the Boltzmann equation. The hadrons
are propagated on classical trajectories and can interact in form of
elastic and inelastic scatterings. Production of new particles via
resonance formation (e.g., $\pi + \pi \rightarrow \rho$) or the decay of
resonances in form of $\Delta \rightarrow N + \pi$. String excitation is
possible for hadron-hadron collisions with $\sqrt{s} > 3$\,GeV but is
negligible in the SIS energy regime considered here.

\subsection{\label{ssec:coarse} Coarse-Grained Contributions} 

For the calculation of thermal emission rates by applying in-medium
spectral functions, one needs to extract the local thermodynamic
properties from the UrQMD simulations. To obtain a phase-space
distribution that is sufficiently smooth in a small volume $\Delta V$ around
each point in space-time, we simulate a large ensemble of events. A grid
of space-time cells with $\Delta x = \Delta y = \Delta z = 0.7 -0.8$\,fm 
and $\Delta t = 0.6$\,fm/$c$   is set up and the energy-momentum 
tensor $T_{\mu\nu}$ as well as the net-baryon four-flow 
$j_{\mu}^{\mathrm{B}}$ are determined in each of these cells as
\begin{equation}
\begin{split}
 T^{\mu\nu}&=\int\dd^{3}p\frac{p^{\mu}p^{\nu}}{p^{0}}f(\vec{x},\vec{p},t)
  =\frac{1}{\Delta V}\left\langle \sum\limits_{i=1}^{N_{h} \in \Delta %
      V} \frac{p^{\mu}_{i}\cdot p^{\nu}_{i}}{p^{0}_{i}}\right\rangle, \\
   j^{\mu}_{\mathrm{B}}&=\int\mathrm{d}^{3}p\frac{p^{\mu}}{p^{0}} f_{\mathrm{B}}(\vec{x},\vec{p},t)
  =\frac{1}{\mathrm{\Delta} V}\left\langle
    \sum\limits_{i=1}^{N_{\mathrm{B}/\bar{\mathrm{B}}} \in \Delta
      V}\pm\frac{p^{\mu}_{i}}{p^{0}_{i}}\right\rangle.
\end{split}
\end{equation} 
The rest frame according to Eckart \cite{Eckart:1940te} can be found by
performing a Lorentz boost such that the baryon flow vanishes in the
cell, i.e. $\vec{j}_{\mathrm{B}}=0$. In the Eckart frame we extract the
energy density $\epsilon$ and baryon density $\rho_{\mathrm{B}}$ as an
input to obtain the temperature $T$ and baryon chemical potential
$\mu_{\mathrm{B}}$ by applying an equation of state. At SIS energies it
is sufficient to use a hadron-resonance gas (HG-EoS)
\cite{Zschiesche:2006rf}. This HG-EoS includes the same degrees of
freedom as the UrQMD model.

It is important to bear in mind that the procedure as described above
assumes local (isotropic) equilibrium in the cell. In macroscopic
descriptions of heavy-ion collisions kinetic and chemical equilibrium is
usually introduced as an ad-hoc assumption. However, we extract the
thermal properties from a transport (i.e.~non-equilibrium) approach
which has effects as viscosity and heat conduction. Although the 
creation of an approximately equilibrated stage is usually considered 
to happen on extremely short time scales after the beginning of the
collision, it is difficult to prove the creation of thermal and
chemical equilibrium explicitly. Previous studies comparing UrQMD
calculations with the results from the statistical thermal model
showed that it might take up to 8-10 fm/$c$ before one can assume the
system to be in approximate kinetic and chemical equilibrium on a global
scale \cite{Bravina:1999dh,Bravina:1999kd}. For practical reasons we 
use the momentum-space anisotropy to characterize to which degree 
the local kinetic equilibrium is constituted in the present study. 
Here, the coarse-grained microscopic transport calculations show 
a significant deviation from equilibrium in form of large pressure 
differences between the longitudinal and transverse components of the 
energy momentum tensor at the beginning of the reaction: During the first 
few fm/$c$ the longitudinal pressure is significantly larger than the 
transverse pressures, which is mainly due to the deposition of high longitudinal
momenta from the colliding nuclei. It has been suggested that in this
case one can determine realistic values for the energy density by using a
generalized equation of state \cite{Florkowski:2010cf,Florkowski:2012pf}
where
\begin{equation} 
\varepsilon_{\text{real}} = \frac{\varepsilon}{r(x)}.
\end{equation} 
For a system with Boltzmann-type pressure anisotropies the relaxation
function $r(x)$ takes the form
\begin{equation} r(x) =\begin{cases} \label{relaxfunc}
\frac{x^{-1/3}}{2}\left(1+\frac{x \artanh \sqrt{1-x}}{\sqrt{1-x}}\right)
\text{for } x \leq 1 \\ \frac{x^{-1/3}}{2}\left(1+\frac{x \arctan
\sqrt{x-1}}{\sqrt{x-1}}\right) \text{for } x \geq 1
    \end{cases},
\end{equation} 
where $x=(P_{\parallel}/P_{\perp})^{3/4}$ denotes the pressure
anisotropy. With this result we can extract meaningful energy
densities from the coarse-grained distributions also in the very early
stage of the reaction, as has already been shown in
\cite{Endres:2014zua}.

With the values of $T$ and $\mu_{\mathrm{B}}$ known for all cells, the
calculation of the thermal dilepton emission is straightforward
\cite{McLerran:1984ay,Rapp:1999ej} and takes the form
\begin{equation}
\begin{split}
\label{rate} \frac{\mathrm{d} N_{ll}}{\mathrm{d}^4x\mathrm{d}^4q} &=
-\frac{\alpha_\mathrm{em}^2 L(M)}{\pi^3 M^2} \; f^{\mathrm{B}}(q \cdot
U;T) \\ & \times 
\im \Pi^{(\text{ret})}_\mathrm{em}(M,
\vec{q};\mu_B,T)
\end{split}
\end{equation} 
with the Bose distribution function $f^{\mathrm{B}}$ and the lepton
phase-space $L(M)$, while $M$ and $q$ denote the invariant mass and 
the momentum of the lepton pair, respectively, and $\alpha_{\mathrm{em}}$ 
is the electromagnetic coupling constant. The relevant physical 
quantity is the retarded electromagnetic current-current correlator
$\Pi^{(\text{ret})}_\mathrm{em}$ which contains all the information on
the medium effects of the spectral function.

Note that equation (\ref{rate}) is derived for the case of full chemical
equilibrium, which requires all chemical potentials of non-conserved
charges to be zero.  Similar to the kinetic anisotropies at the
beginning of the reaction, also deviations from the chemical equilibrium
composition of the system appear at the early stages of transport
simulations. Especially an overpopulation of pions is observed here,
which dominates the evolution for a significant period of time
\cite{Bandyopadhyay:1993qj}. The appearance of a finite pion chemical
potential $\mu_{\pi}$ was explained by the large initial production of
pions and their long relaxation time
\cite{Bebie:1991ij,Kataja:1990tp}. This is of particular importance for
the present study since a finite chemical potential $\mu_{\pi}$ was
considered to have a significant influence on the thermal dilepton
emission rates by its influence on the $\pi$-$\rho$ interactions
\cite{Baier:1997xc,Baier:1997td}. For the present calculations, we
extract the pion chemical potential in Boltzmann approximation for each
cell and consider its effect on the thermal dilepton emission (i) in
form of its direct influence on the spectral functions and (ii) as
additional fugacity factor, which will show up in equation (\ref{rate})
for the chemical non-equilibrium case.

In the following the different thermal contributions which are
considered for the present study are discussed.

\subsubsection{\label{sssec:rho} Thermal $\rho$ and $\omega$ Emission}

Several approaches for the description of in-medium effects on vector
mesons exist. However, a full description of the different effects that
influence the spectral shape in a hot and dense medium are highly
non-trivial, and there are only a few calculations that include both the
effects of finite temperature and density. E.g.~in
Ref.~\cite{Eletsky:2001bb} the in-medium spectral functions were 
determined using empirical scattering amplitudes. However, this spectral
function is calculated in low-density approximation for a weakly
interacting pion-nucleon gas and tested only up to densities of
$2\rho_{\mathrm{0}}$. The application of this approach is therefore
questionable for the situation at SIS energies where very high net 
baryon densities are reached.

In the present work a calculation from hadronic many-body theory
\cite{Rapp:1997fs,Urban:1998eg,Rapp:1999us,Rapp:1999qu} is applied,
which has proven to successfully describe the dilepton spectra at SPS
and RHIC energies \cite{vanHees:2007th, Rapp:2000pe}. In the medium,
three different contributions to the self-energy of the $\rho$ are taken
into account here. These are
\begin{enumerate}[(a)]
\item the modification of the pion cloud of the $\rho$ meson by
  particle-hole and $\Delta$-hole excitations in the medium,
\item scattering with mesons ($M = \pi,K,\bar{K},\rho$) and
\item scattering with the most abundant baryon resonances
  ($B = N,\Delta_{1232},N^{*}_{1440},\dots$).
\end{enumerate} The in-medium propagator consequently takes the form
\begin{equation}
D_{\rho}=\frac{1}{M^{2}-m_{\rho}^{2}-\Sigma_{\rho\pi\pi}-\Sigma_{\rho
M}-\Sigma_{\rho B}}.
\end{equation} 
The $\omega$ spectral function \cite{Rapp:2000pe} is similarly
constructed. However, here the situation is more complicated as the
$\omega$ is basically a three-pion state, and the vacuum self energy is
given by decays into $\rho \pi$ or $3\pi$. The full in-medium propagator
reads
\begin{equation}
\begin{split} D_{\omega} =
&[M^{2}-m_{\omega}^{2}+\mathrm{i}\,m_{\omega}\left(\Gamma_{3\pi}+\Gamma_{\rho\pi}
+\Gamma_{\omega\pi\rightarrow\pi\pi}\right) \\ &\,-\Sigma_{\omega\pi
b_{1}}-\Sigma_{\omega B}]^{-1}.
\end{split}
\end{equation} 
It includes the following contributions:
\begin{enumerate}[(a)]
\item $\omega \rightarrow \rho \pi$ decays including the corrections
  from the medium-modified $\rho$ spectral function,
\item the direct $\omega \rightarrow 3\pi$ decays,
\item $\omega \pi \rightarrow \pi \pi$ inelastic scattering,
\item $\omega \pi \rightarrow b_{1}$ resonance scattering and
\item $\omega N \rightarrow N_{(1520)},\,N_{(1650)}$ resonance
  scattering at an effective nucleon density $\rho_{\mathrm{eff}}$.
\end{enumerate} 
To take into account the off-equilibrium of the pions an additional
fugacity factor is introduced in equation (\ref{rate}), as already
explained above. In one-loop approximation, for the thermal $\rho$ and
$\omega$ emission one obtains an additional overall fugacity factor
\cite{Rapp:1999ej}
\begin{equation} z^{n}_{\pi}=e^{n \mu_{\pi}/T}.
\end{equation} 
The exponent $n$ hereby takes the value 2 in case of the $\rho$ emission
and 3 for the $\omega$ contribution \cite{vanHees:2006ng}. The final
yield is then calculated according to
\begin{equation} \frac{\mathrm{d}N_{e^{+}e^{-}}}{\mathrm{d}M}=\Delta
x^{4}\int\frac{\mathrm{d}^{3}p}{p_{0}}\frac{\mathrm{d}N_{e^{+}e^{-}}}{\mathrm{d}^{4}x\mathrm{d}^{4}p}z^{n}_{\pi}.
\end{equation} 
Note that the four-volume $\Delta x^{4} =\Delta t \Delta V$ of the cell
is an invariant quantity and therefore the same in all reference frames.

As it is not reasonable or possible to calculate thermal dilepton emission 
for all cells (e.g., due to low temperature) we also include a non-thermal 
(``freeze-out'') contribution for the $\rho$ and $\omega$ meson. Those 
dileptons are directly extracted from the UrQMD calculations (for details 
see section \ref{sssec:transrho}).

\subsubsection{\label{sssec:multipi} Multi-$\pi$ Contribution} 

Although it is known from previous theoretical investigations at higher
collision energies that the multi-pion contribution plays a significant role only
for masses greater than 1\,GeV/$c^{2}$, which is mostly beyond the
kinematic limit of dilepton production at SIS energies, we include this
contribution for completeness. Here the same description as developed in
\cite{vanHees:2006ng, vanHees:2007th} is applied which uses chiral
reduction techniques for the case of the chiral limit
\cite{Dey:1990ba}. This leads to a chiral mixing of the isovector part
of the vector and axial-vector correlators. The isovector-vector current
correlation function takes the form
\begin{equation}
\begin{split}
  \label{vamix} 
  \Pi_{V}(p) = &
  (1-\varepsilon)z_{\pi}^{4}\Pi^{\text{vac}}_{\text{V},4\pi} +
  \frac{\varepsilon}{2}z^{3}_{\pi} \Pi^{\text{vac}}_{\text{A},3\pi} \\
  &+
  \frac{\varepsilon}{2}(z^{4}_{\pi}+z^{5}_{\pi})\Pi^{\text{vac}}_{\text{A},5\pi},
\end{split}
\end{equation} 
where $\hat{\varepsilon}$ denotes the mixing parameter which depends on
temperature, critical temperature and the pion chemical potential. The
result is an admixture of three-pion and five-pion axial-vector pieces
on the vector four-pion part. Once again, as in case of the $\rho$ and
$\omega$ spectral functions, the effect of a finite pion chemical
potential also enters in form of a fugacity factor $z_{\pi}^{n}$ with
$n=3,4,5$ denoting the pion multiplicity of the corresponding
state. Note that the two-pion and the three-pion piece corresponding to
the decay $a_{1} \rightarrow \rho + \pi$ are already included in the
$\rho$ spectral function and are therefore not considered for the
multi-pion contribution.

\subsection{\label{ssec:shining} Transport Contributions}

\subsubsection{\label{ssec:pietadal} $\pi$ and $\eta$ Dalitz Decay} 

The $\pi^{0}$ and the $\eta$ pseudo-scalar mesons can both decay into a
lepton pair via the Dalitz decays
\begin{equation} \pi^{0} \rightarrow \gamma\,l^{+}l^{-}, \ \ \eta
\rightarrow \gamma\,l^{+}l^{-}.
\end{equation} 
Following the scheme presented in \cite{Schmidt:2008hm}, we treat this
decay as a two-step process: First the decay of the pseudo-scalar meson
into a photon $\gamma$ and a virtual photon $\gamma^{*}$ and the
following electromagnetic conversion of the $\gamma^{*}$ into a lepton
pair \cite{Landsberg:1986fd}. The width of the meson decaying into a
photon and a virtual photon can be related to the radiative decay
width. The full expression then takes the form
\begin{equation}
\begin{split}
\label{Dalitz} 
\frac{\dd \Gamma_{M\rightarrow \gamma l^{+}l^{-}}}{\dd M^{2}} =
&2 \Gamma_{M\rightarrow 2 \gamma}\left(1 -
\frac{M^{2}}{m_{M}^{2}}\right)^{3} \\ & \times
|F_{M\gamma\gamma^{*}}(M^{2})|^{2}\frac{\alpha_{em}}{3\pi M^{2}} L(M).
\end{split}
\end{equation} 
With the lepton phase space $L(M)$ and the form factor
$F_{P\gamma\gamma^{*}}(M^{2})$. Here we use the form factors as obtained
from fits to experimental data, which are in very good agreement with
the values as predicted by the Vector Dominance Model
\cite{Landsberg:1986fd}:
\begin{equation}
\begin{split} F_{\pi^{0}\gamma\gamma^{*}}(M^{2})&=1+b_{\pi^{0}}M^{2}, \\
F_{\eta\gamma\gamma^{*}}(M^{2})&=\left(1-\frac{M^{2}}{\Lambda^{2}_{\eta}}\right)^{-2},
\end{split}
\end{equation} 
with $b_{\pi^{0}}=5.5$\,GeV$^{-2}$ and $\Lambda_{\eta}=0.72$\,GeV. Note that for this contribution, we only consider those particles from
the final freeze-out of the calculation, neglecting all $\pi$ and $\eta$
which are reabsorbed during the evolution of the collision.

\subsubsection{\label{ssec:shphi} $\phi$ Direct Decay} 

The width for the direct decays of the vector mesons can be determined
via \cite{Li:1996mi}
\begin{equation}
\label{dirwidth} \Gamma_{V \rightarrow ll}(M) = \frac{\Gamma_{V
\rightarrow ll}(m_{V})}{m_{V}} \frac{m_{V}^{4}}{M^{3}} L(M).
\end{equation} 
For the branching ratio at the meson pole mass $m_{V}$ we use the value
from the particle data group,
$\Gamma_{\phi\rightarrow ee}(m_{V}) / \Gamma_{\text{tot}}= 2.95 \cdot 10^{-4}$
\cite{Agashe:2014kda}.

\begin{figure*}
\includegraphics[width=1.0\columnwidth]{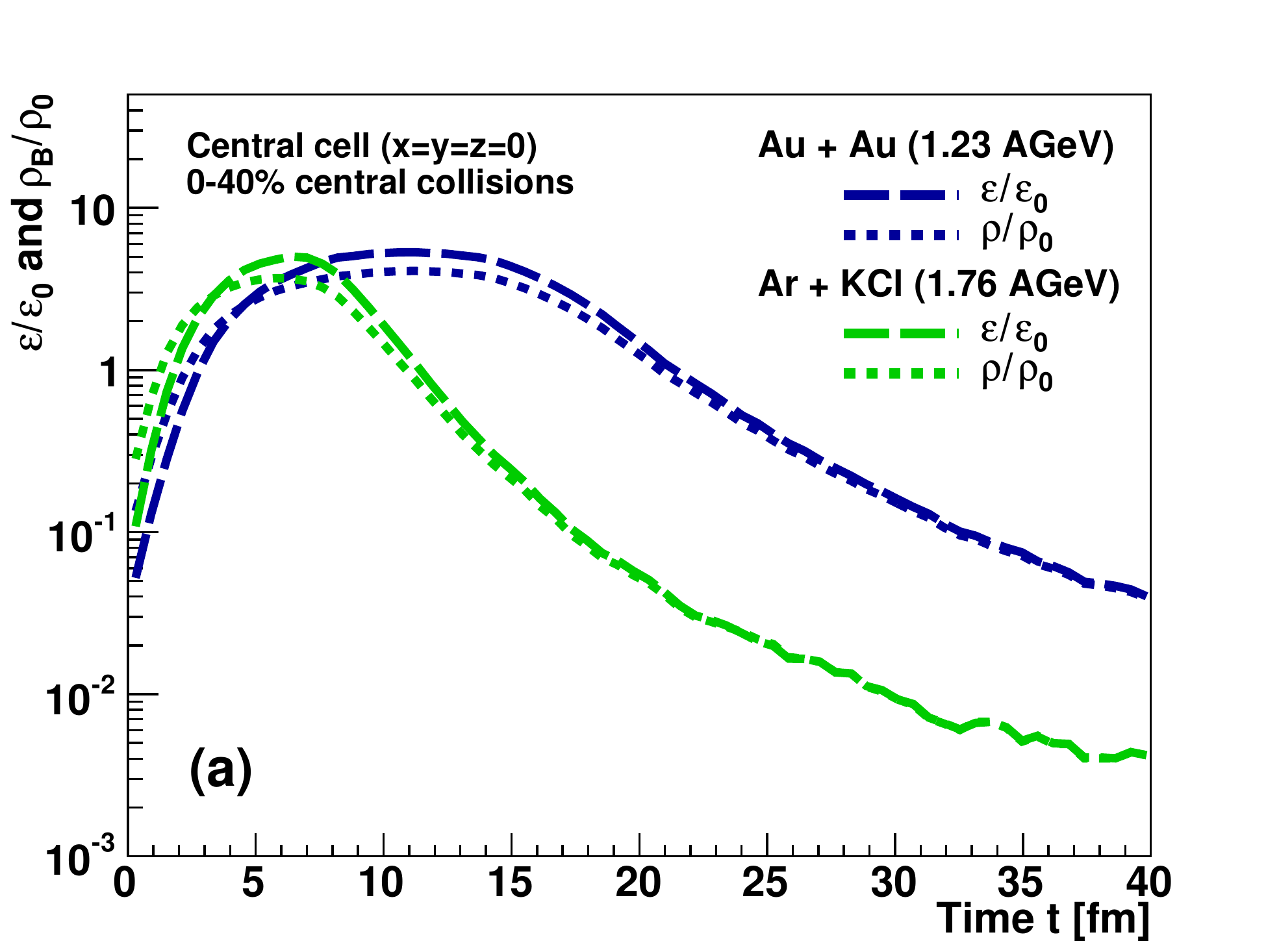}
\includegraphics[width=1.0\columnwidth]{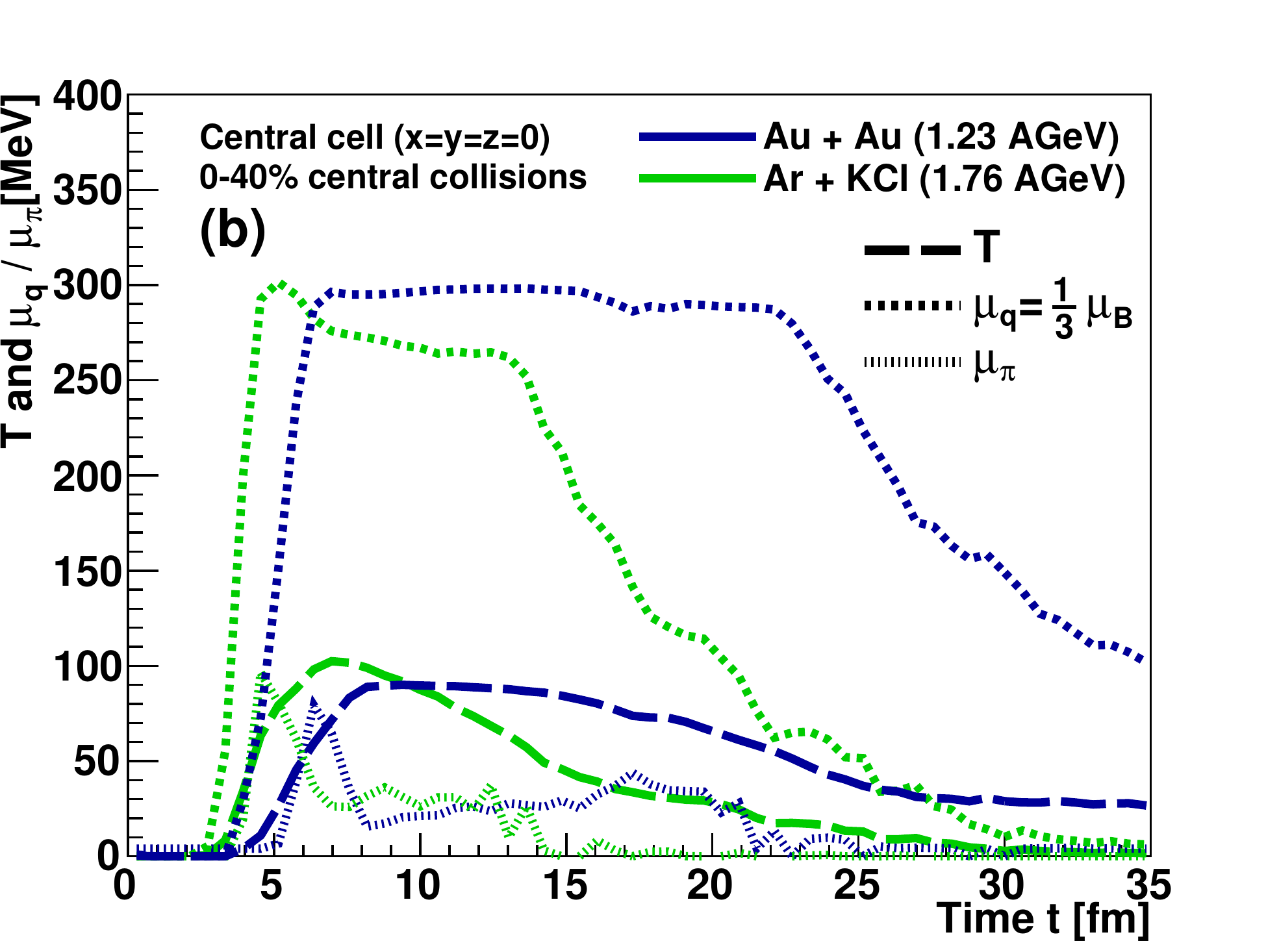}
\caption{(Color online) Time evolution of the energy and baryon
  densities $\varepsilon$ and $\rho$ in units of the according
  ground-state densities (a) and for temperature $T$ as well as the
  baryon and pion chemical potential, $\mu_{\mathrm{B}}$ respectively $\mu_{\pi}$
  (b) for the central cell (i.e., at $x=y=z=0$). The results as obtained
  via coarse-graining of microscopic transport calculations are shown
  for Ar+KCl collisions at $E_{\mathrm{lab}}=1.76$\,$A$GeV and for Au+Au
  reactions at 1.23\,$A$GeV.}
\label{densev}
\end{figure*}
An important difference compared to the procedure for the long-lived
pseudo-scalar mesons is the assumption that the $\phi$ mesons
continuously emit dileptons over their lifetime
\cite{Heinz:1991fn}. That means, we track the time of production and the
decay (or absorption) of the $\phi$ and integrate over the corresponding
lifetime in the particle's rest frame.
\begin{equation}\label{phishining} \frac{\mathrm{d}N_{ll}}{\mathrm{d}M}
=\frac{\Delta N_{ll}}{\Delta M}= \sum _{i=1}^{N _{\Delta M}}\sum
_{j=1}^{N _{\rho}} \int_{t_{i}}^{t_{f}} \frac{\mathrm{d}t}{\gamma_{V}}
\frac{\Gamma_{V\rightarrow ll}(M)}{\Delta M}.
\end{equation} 
The factor $\gamma_{V}$ is here introduced to account for the
relativistic time dilation in the computational frame compared to the
vector mesons rest frame.
 
This ``shining approach'' is identical with the
assumption of simply having one additional dilepton from each $\phi$
(weighted with the according branching ratio) if there is no
absorption. But inside a dense medium, there is a significant chance
that the particle will not decay but suffer an inelastic
collision. Therefore the probability for the decay into a dilepton is
reduced, and the shining as described above accounts for this effect
\cite{Ernst:1997yy}.

\subsubsection{\label{sssec:transrho} Non-thermal $\rho$ and $\omega$}

There are two situations, where a thermal calculation of the dilepton
emission from spectral functions becomes difficult or even
unreasonable. Firstly, for cells with no baryon content, where the
Eckart definition of the rest-frame does not apply (those cells are
found in the late stage of the reaction in very peripheral cells)
and secondly for cells where the temperature is found to be below
50\,MeV. In the latter case, when going to such usually low-density
cells, the determination of $T$ and $\mu_{\mathrm{B}}$ becomes less accurate and
one will necessarily come to the point where the assumption of a
thermalized system in the cell becomes unreliable. Consequently, 
we do not assume any thermal dilepton emission
here. This procedure is in line with the findings of thermal-model
studies \cite{Cleymans:1998yb}, where it was shown that the freeze-out
temperature in heavy-ion collisions at $E_{\mathrm{lab}} =1-2$\,$A$GeV
is around 50\,MeV. This also indicates that it is neither necessary nor
suggestive to assume thermalization of the system at lower temperatures.

However, one has to consider that dilepton emission from $\rho$ and
$\omega$ mesons is of course also possible in the cells for which one
of the conditions mentioned above pertains. As the macroscopic picture
is questionable here, we apply a similar procedure as for the
$\phi$ (described in section \ref{ssec:shphi}) to extract the dilepton
emission from the microscopic simulation. The width for the direct
decays of the vector mesons can likewise be determined via equation
(\ref{dirwidth}). For the branching ratio at the meson-pole mass we use
the values from the particle data group, i.e.
$\Gamma_{\rho\rightarrow ee} / \Gamma_{\text{tot}}= 4.72 \cdot 10^{-5}$
and
$\Gamma_{\omega\rightarrow ee} / \Gamma_{\text{tot}}= 7.28 \cdot
10^{-5}$
\cite{Agashe:2014kda}. However, the $\omega$ can not only decay
into a lepton pair directly, but also via Dalitz conversion into a pseudo-scalar
meson and a dilepton. As in the case of the $\pi$ and the $\eta$
equation (\ref{Dalitz}) applies here. Only the form factor is different
and takes the form
\begin{equation}
|F_{\omega}(M^{2})|^{2}=\frac{\Lambda_{\omega}^{2}\left(\Lambda_{\omega}^{2}
+ \gamma_{\omega}^{2}\right)}{\left(\Lambda_{\omega}^{2} - M^{2}\right)+
\Lambda_{\omega}^{2} \gamma_{\omega}^{2}}
\end{equation} 
with the parameters $\Lambda_{\omega} = 0.65$\,GeV and
$\gamma_{\omega} = 0.04$\,GeV \cite{Bratkovskaya:1999mr,Schmidt:2008hm}.

\begin{figure}
\includegraphics[width=1.0\columnwidth]{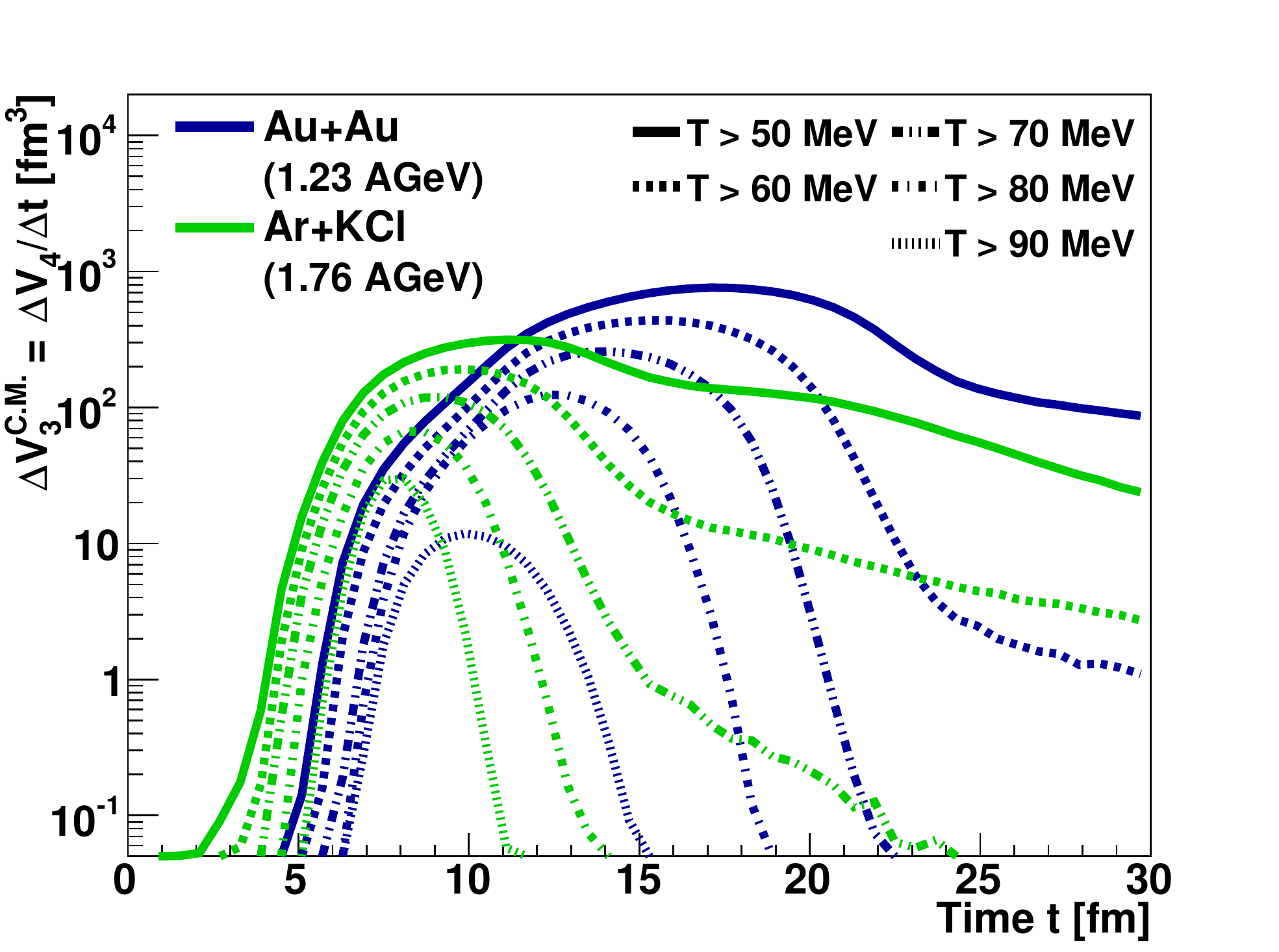}
\caption{(Color online) Time evolution of the thermal volume $V_{3}^{\mathrm{C.M.}}$
as seen from  the center-of-momentum frame of the collision. This is equal to the thermal four-voulume $V_{4}$ for each timestep divided by the length of $\Delta t$, which is 0.6\,fm/$c$ here. The results are shown for Ar+KCl collisions at
$E_{\mathrm{lab}}=1.76$\,$A$GeV and for Au+Au reactions at 1.23\,$A$GeV 
as obtained via coarse-graining of microscopic transport calculations. 
The results are plotted for different temperatures.}
\label{4volev}
\end{figure}
\begin{figure*}
\includegraphics[width=1.0\columnwidth]{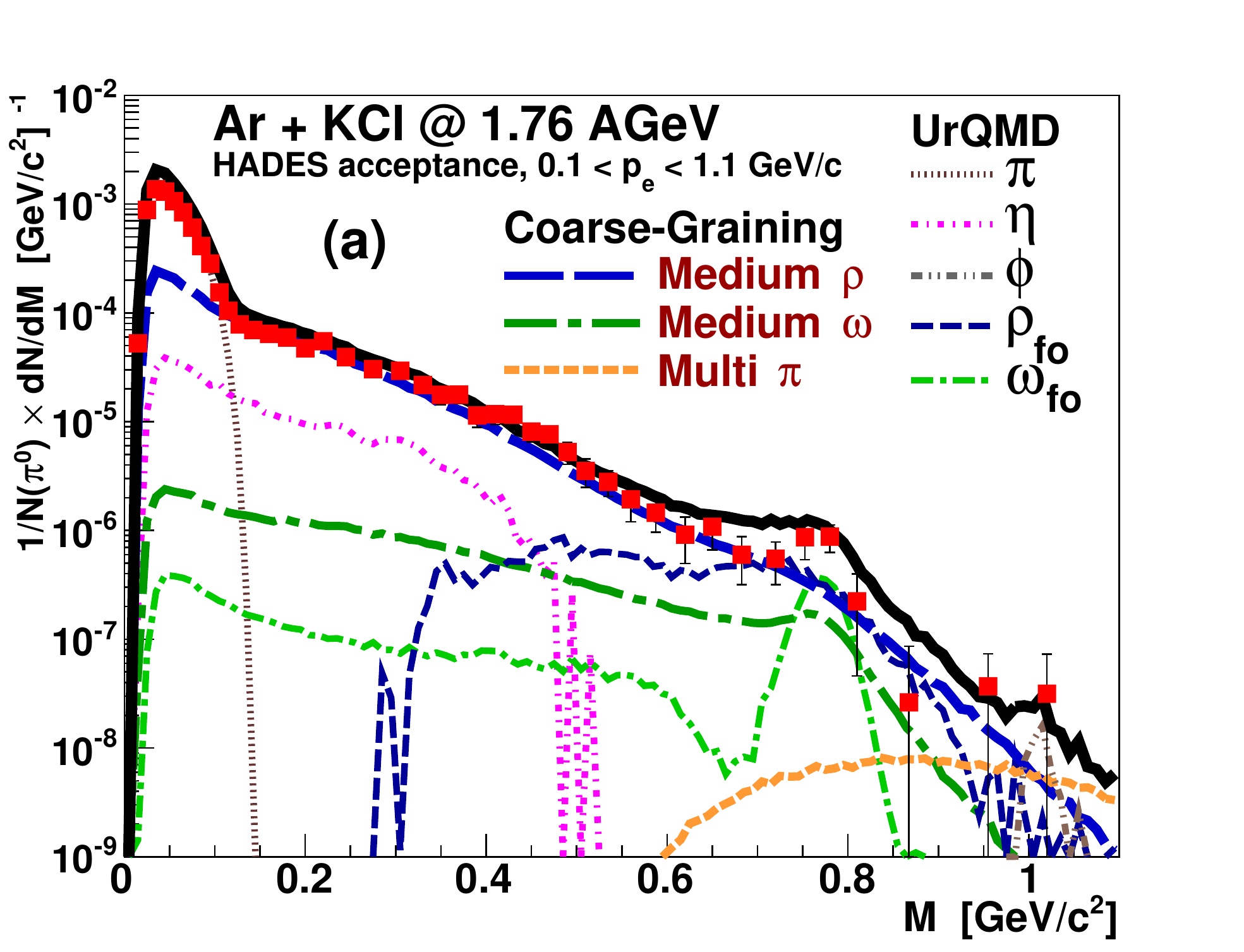}
\includegraphics[width=1.0\columnwidth]{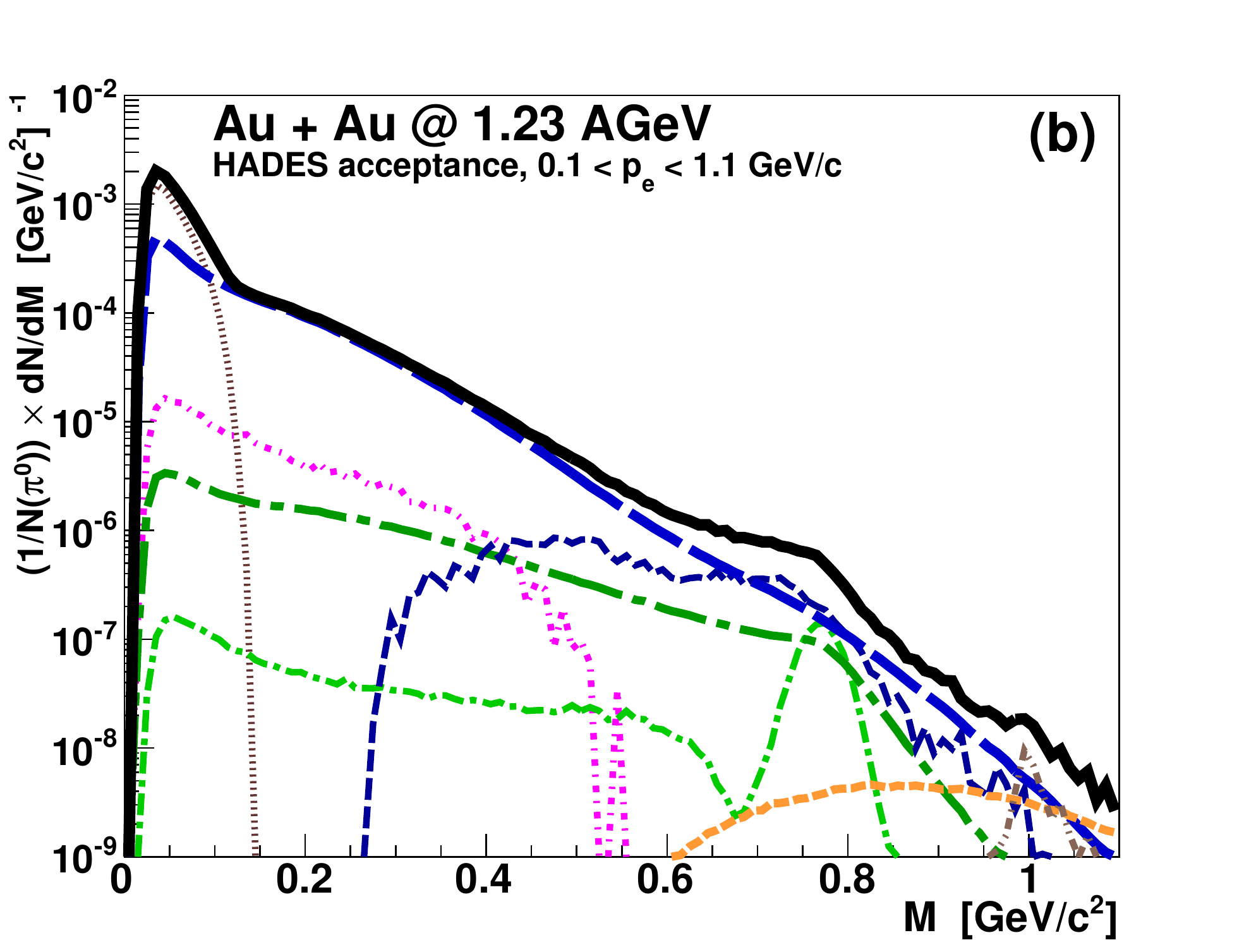}
\caption{\label{invmass} (Color online) Invariant mass spectra of the
  dielectron yield for Ar+KCl collisions at $E_{\mathrm{lab}}=1.76$\,$A$GeV (a)
  and for Au+Au at $E_{\mathrm{lab}}=1.23$\,$A$GeV (b). The results are
  normalized to the average total number of $\pi^{0}$ per event and
  shown within the HADES acceptance. The results for Ar+KCl are compared
  to the experimental data from the HADES Collaboration
  \cite{Agakishiev:2011vf}.}
\label{invmassCaAu}
\end{figure*}
Similarly to the procedure for the $\phi$ meson, a continuous emission
of dileptons from the $\rho$ and $\omega$ is assumed in these special cases.
However, as we consider space-time cells with a definite length of the time-steps
$\Delta t$, the dilepton rate is multiplied with this time instead of
the actual particle's lifetime within the cascade simulation. This is
done to guarantee consistency, avoid double counting and strictly
distinguish between cells with thermal and non-thermal
emission. Consequently, here equation (\ref{phishining}) takes the form
\begin{equation}\label{shining} \frac{\mathrm{d}N_{ll}}{\mathrm{d}M}
=\frac{\Delta N_{ll}}{\Delta M}= \sum _{i=1}^{N _{\Delta M}}\sum
_{j=1}^{N _{V}}\frac{\Delta t}{\gamma_{V} } \frac{\Gamma_{V \rightarrow
ll}(M)}{\Delta M},
\end{equation}
and is applied for each time step $\Delta t$.
\section{\label{sec:2} Results}
For the results presented here we used calculations with an ensemble of
1000 UrQMD events. However, several runs using different UrQMD events 
as input had to be performed to obtain enough statistics especially for
the non-thermal $\rho$ and $\omega$ contributions. Note that in case of the experimental Ar+KCl reaction
we simulated the collision of two calcium ions instead, as this makes
the calculation easier for symmetry reasons. Effectively it is the same
as the Ar+K or Ar+Cl reactions that were measured in the experiment and
the size of the system remains identical. To simulate the correct impact
parameter distribution, we made a Woods-Saxon type fit to the HADES
trigger conditions for Ar+KCl \cite{Agakishiev:2010rs} and Au+Au
\cite{GalatyukPC}. In both cases this approximately corresponds to a
selection of the 0-40\% most central collisions. The number of $\pi^{0}$
per event, which will be important for the normalization of the dilepton
spectra, are found to agree well with the HADES measurement for Ar+KCl
reactions. Here the HADES collaboration measured
$N_{\pi^{0}}^{\mathrm{exp}} = 3.5$ where we find
$N_{\pi^{0}}^{\mathrm{sim}} \approx 3.9$, i.e. the deviation is only
12\%. For the larger Au+Au system a number
$N_{\pi^{0}}^{\mathrm{sim}} \approx 8.0$ results from the events
generated with the UrQMD model. Note that for reasons of self-consistency
we normalize the dilepton spectra with the UrQMD $\pi^{0}$ yield, not
the experimental one.

The final dilepton results were filtered with the HADES acceptance
filter \cite{HADESwiki}, and momentum cuts were applied to compare the
simulations with the experimental results. As only very preliminary
results and no filters are available for the Au+Au reactions at
1.23\,$A$GeV, we used the same filter as for p+p and p+n reactions at
1.25\,$A$GeV which should be quite close to the final acceptance
\cite{GalatyukPC}.

In case of the DLS Ca+Ca spectrum, version 4.1 of the DLS acceptance
filter \cite{DLSfil} is used. Furthermore, an RMS smearing of 10\% is
applied to account for the detector resolution. For this reaction we
used a minimum-bias simulation of Ca+Ca events, because impact-parameter
distributions are not available for DLS. Here the final invariant-mass
spectrum is normalized to the total cross-section of a Ca+Ca reaction.

\subsection{\label{ssec:ReacDyn} Reaction Dynamics}
\begin{figure*}
\includegraphics[width=1.0\columnwidth]{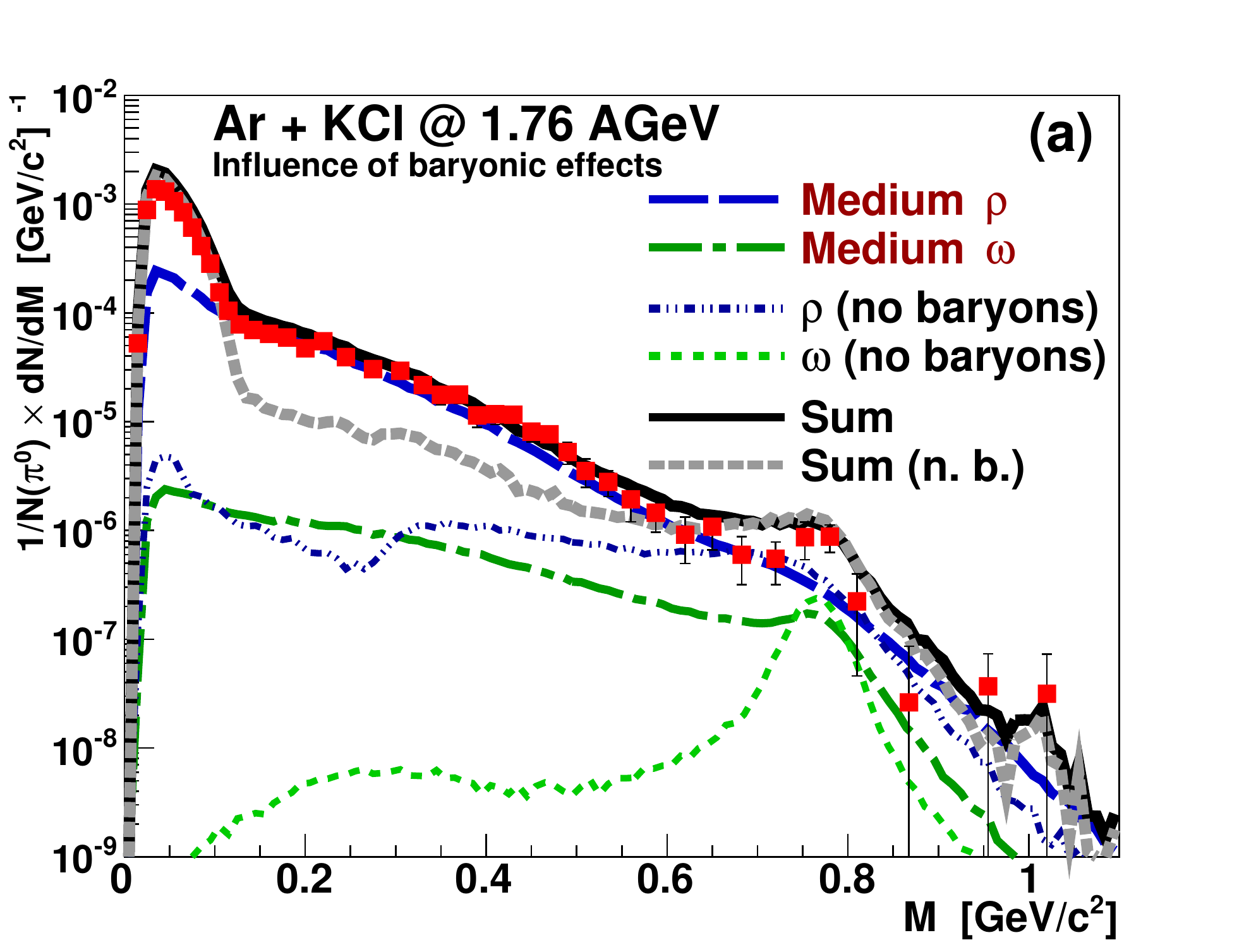}
\includegraphics[width=1.0\columnwidth]{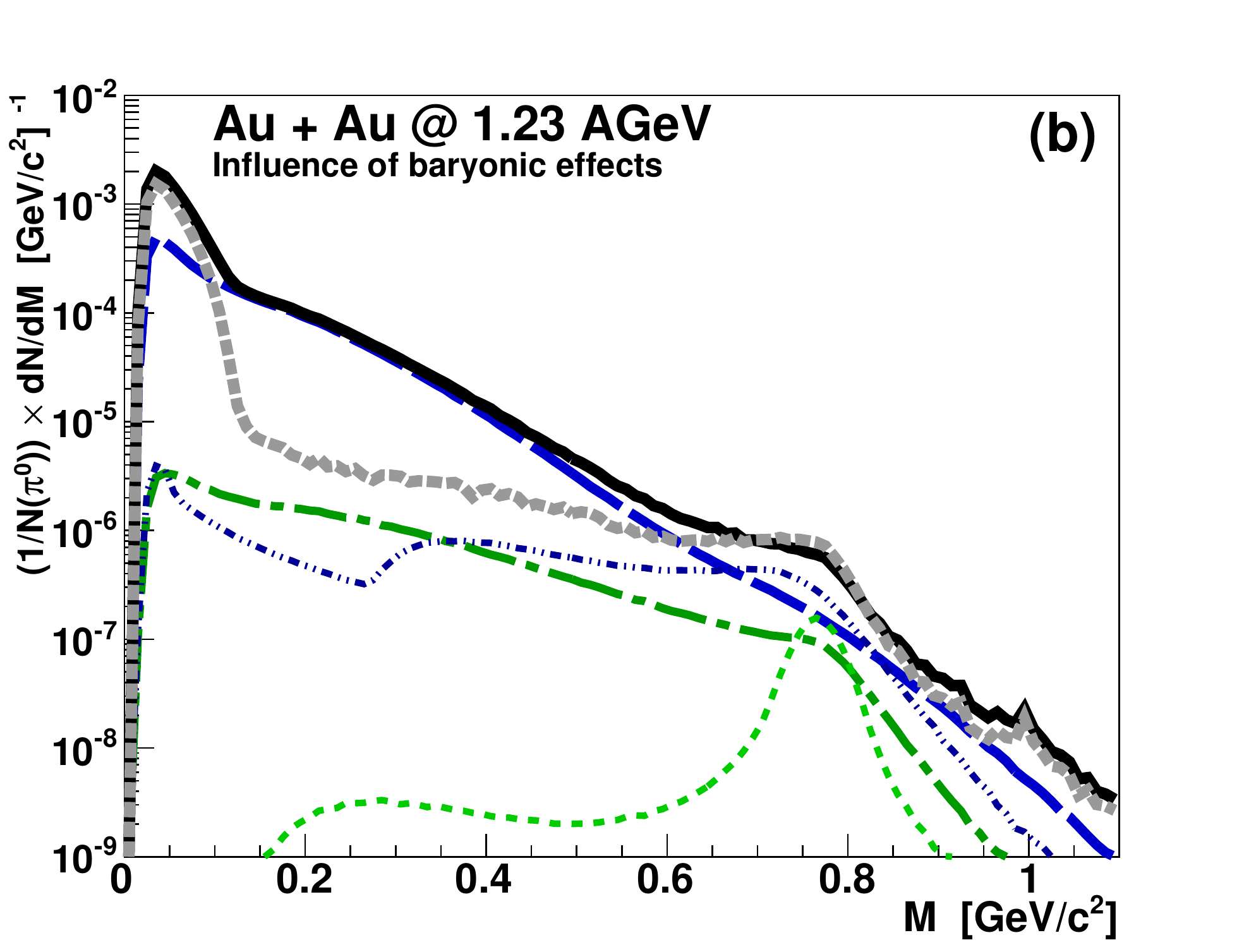}
\caption{\label{nobar} (Color online) Comparison of invariant mass
  spectra with the full spectral function and for the case of no
  baryonic effects (i.e. for $\rho_{\mathrm{eff}}=0$). As in 
  Fig.~\ref{invmass}, the dielectron yields for Ar+KCl collisions at
  $E_{\text{lab}}=1.76$\,$A$GeV (a) and for Au+Au at
  $E_{\mathrm{lab}}=1.23$\,$A$GeV (b) are shown within HADES acceptance
  and normalized to the average number of produced $\pi^{0}$. Note that
  the UrQMD contributions are included in the sum, but the different
  single yields are not shown explicitly for reasons of lucidity.}
\label{vaccomp}
\end{figure*}
The main difference between the two heavy-ion reactions considered here (as measured by the HADES Collaboration) is the
size of the colliding nuclei. Therefore, it is interesting to first have
a look at the evolution of the reaction for both systems. In
Fig.~\ref{densev} the evolution of the baryon and energy density (a) as
well as the evolution of temperature and the chemical potentials (b) are
shown. The maximum density values in the central cell of
the grid, i.e., in the center of the collision, reach similar values up
to roughly 3-6 times ground-state baryon density $\rho_{0}$ and energy
density $\varepsilon_{0}$ for both reactions. In case of the larger system
(Au+Au) a plateau develops for a duration of more than 10 fm/$c$, while
for Ar+KCl (respectively Ca+Ca in our simulations) the densities drop
off rather quickly after reaching the maximum. Note that the values for
the energy density $\varepsilon$ shown in Fig.~\ref{densev}\,(a) are
corrected for the pressure anisotropy at the beginning of the collision.
 
With regard to the evolution of temperature and baryochemical potential
in Fig.~\ref{densev}\,(b), we find that again similar top values are
obtained. In both reactions, Ar+KCl and Au+Au, we get peak values of
$T \approx 100$ MeV and $\mu_{\mathrm{B}} \approx 900$ MeV. (Note that
the quark chemical potential $\mu_{q}=\frac{1}{3} \mu_{\mathrm{B}}$ is
shown instead of the baryon chemical potential.) Especially for the
Au+Au reaction the baryon chemical potential shows a more prominent
plateau than the baryon density. However, note that $T$ and
$\mu_{\mathrm{B}}$ depend on $\varepsilon$ and $\rho_{\mathrm{B}}$
non-linearly via the EoS. For the collisions at SIS energies studied
here, $\mu_{\mathrm{B}}$ rapidly rises to values very close to the
nucleon mass - but once it reaches such a high level it exhibits a much
less significant rise in spite of a further increase of
$\rho_{\mathrm{B}}$. This is a consequence of the Fermi statistics which
takes effect in this case. Furthermore, it is clear that the values of
$\mu_{\mathrm{B}}$ are much higher here and show a different evolution
than in our recent study for top SPS energy \cite{Endres:2014zua}. In
case of Au+Au collisions the central cell stays for approximately
20\,fm/$c$ in a stage with extremely high baryochemical potential, so
that any baryon-driven effect on the dilepton spectra should be clearly
visible for this system. Similar findings are also true for Ar+KCl
reactions, but with significantly shorter lifetime. The maximum
temperature is approximately 10 MeV higher in the latter reaction due to
the slightly higher collision energy.

The pion chemical potential rises up to values of around 100 MeV and
then equally quickly drops to values around 20-50 MeV for the rest of
the reaction. The peak at the beginning can be explained with the
non-equilibrium nature of the cascade, where a large number of pions is
produced rapidly at the very beginning before the system has time to
equilibrate.

The corresponding evolution of the thermal four-volume at each time
step, divided by its duration $\Delta t$, as obtained for both systems
from the coarse-grained microscopic transport calculations is shown in
Fig.~\ref{4volev}. This quantity is the spatial volume for the different
temperature ranges as seen from the center-of-momentum frame of the
collision. One observes that the hotter cells created during the
evolution of the system (for temperatures above 80 respectively 90\,MeV)
are present only for comparatively short time spans, and their number is
significantly smaller than the volume of cells with lower
temperatures. While we only show the evolution of the first 30\,fm/$c$
in Fig.~\ref{4volev} for the sake of clarity, we mention that even at
the end of the simulation after 70\,fm/c one still finds a few cells at
temperatures above 50 MeV. Note, however, that the dilepton emission
from these late stage cells is marginal and insignificant with respect to
the overall time-integrated dilepton spectra.

\subsection{\label{ssec:DilSpec} Dilepton Spectra}
\subsubsection{HADES Results}
\begin{figure*}
\includegraphics[width=1.0\columnwidth]{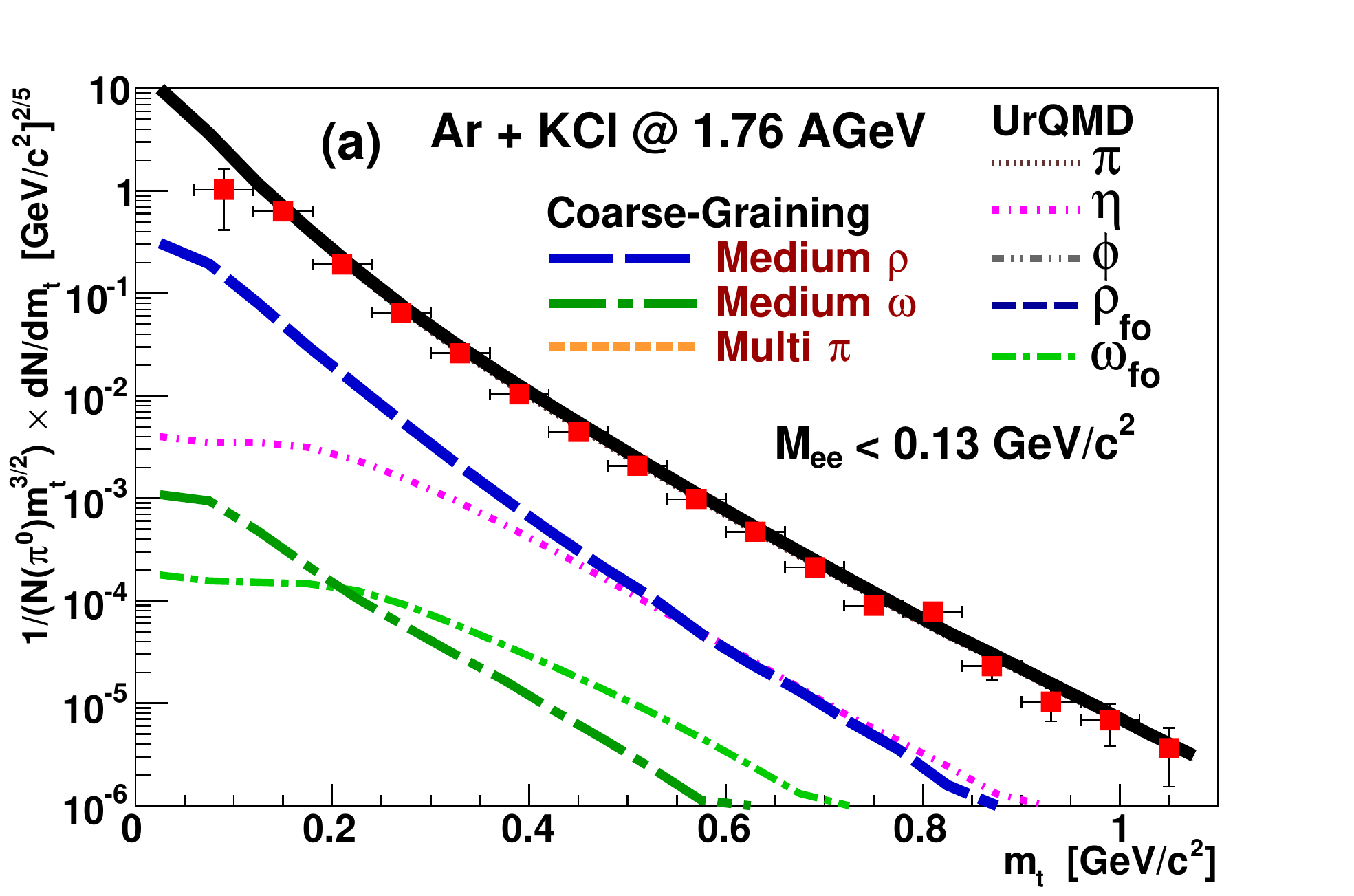}
\includegraphics[width=1.0\columnwidth]{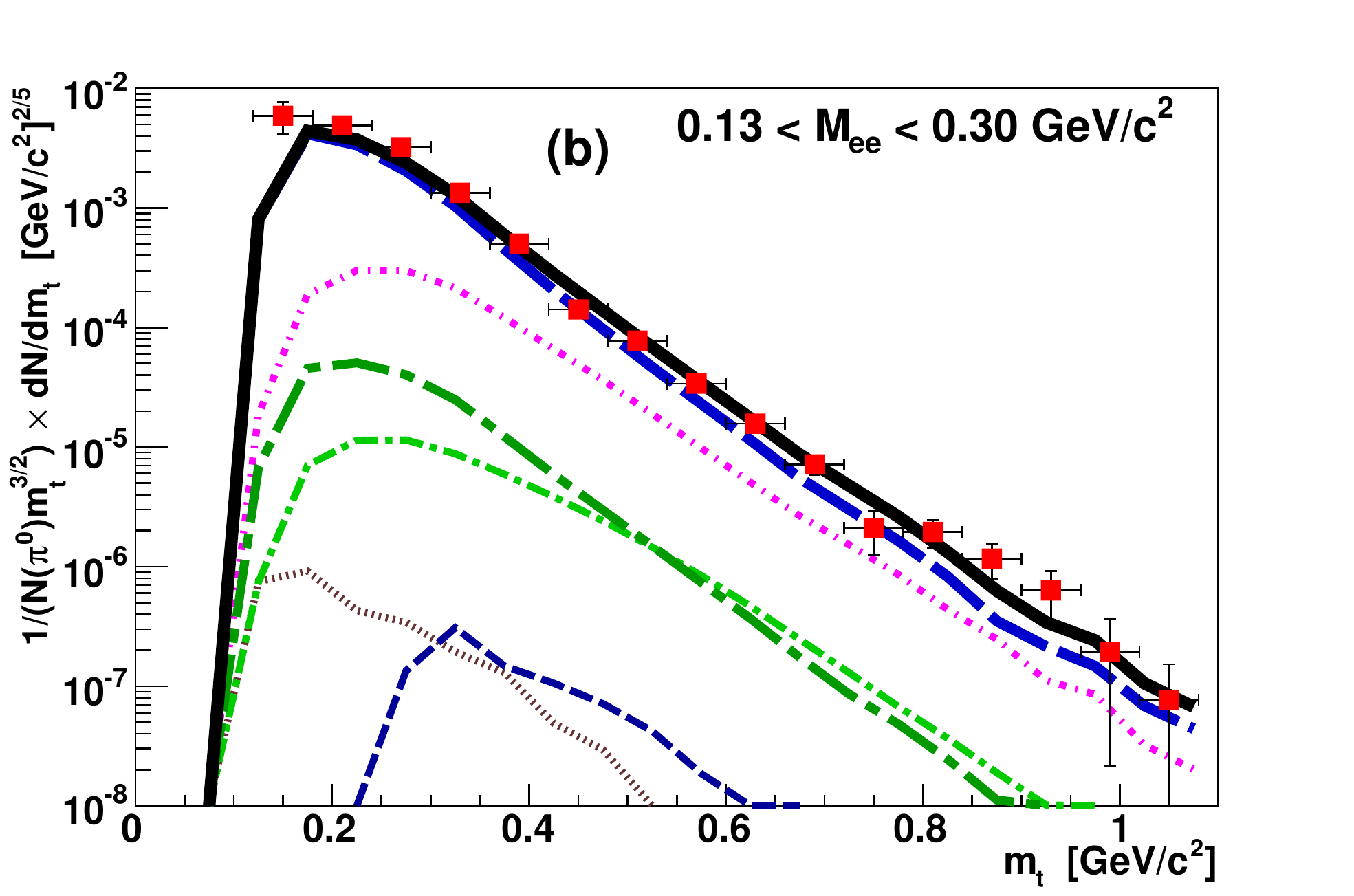}
\\
\includegraphics[width=1.0\columnwidth]{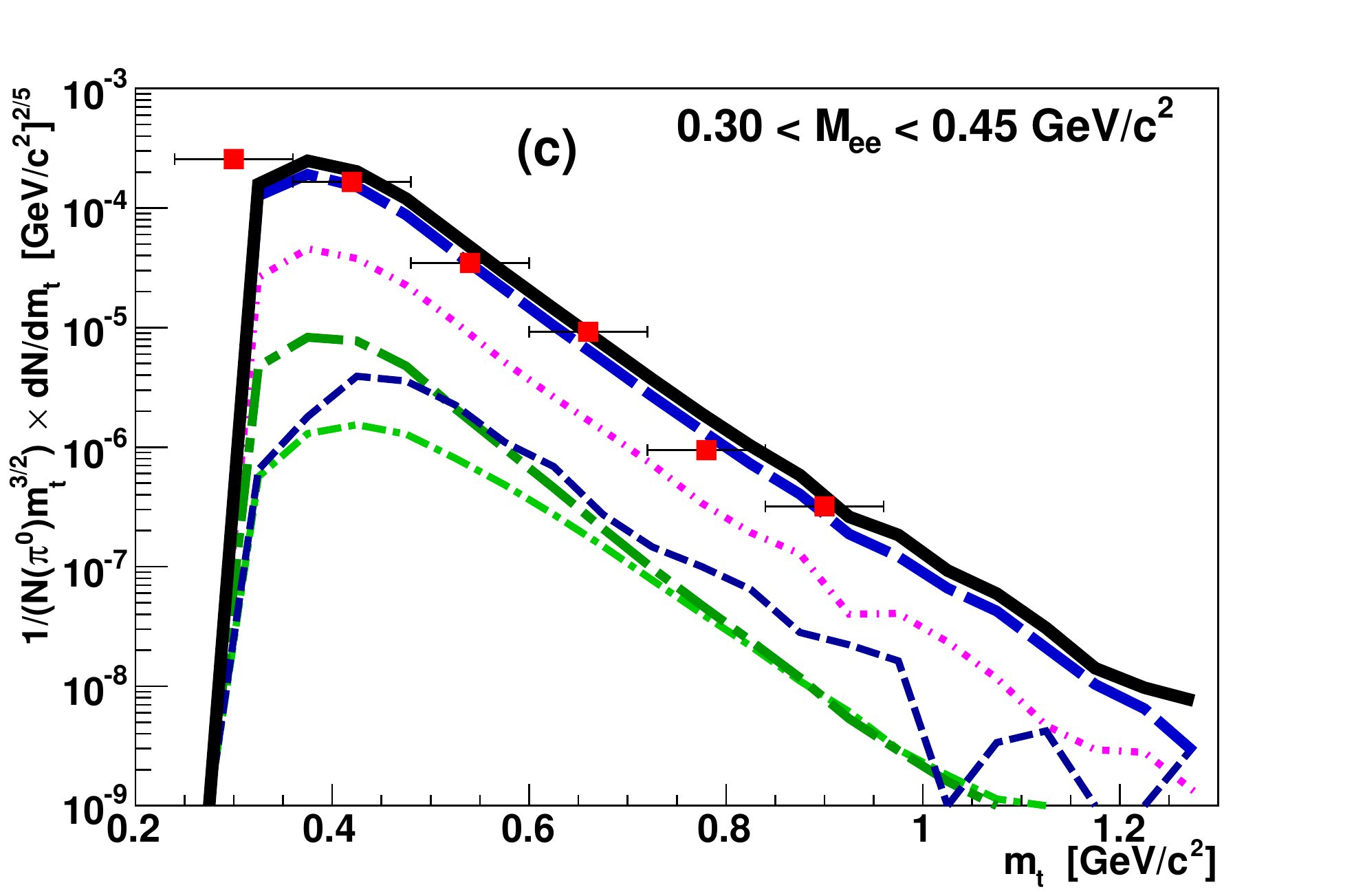}
\includegraphics[width=1.0\columnwidth]{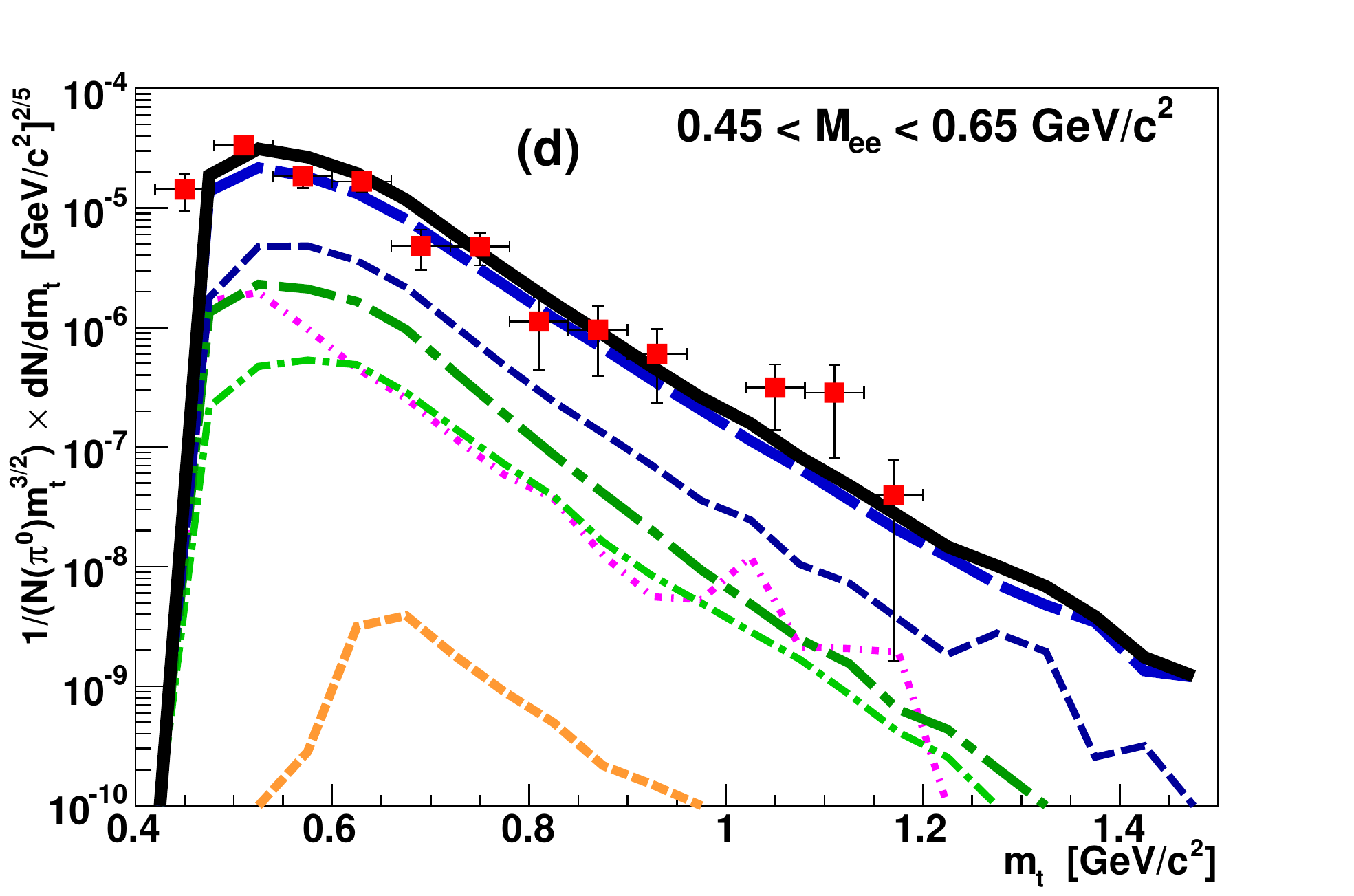}
\\
\includegraphics[width=1.0\columnwidth]{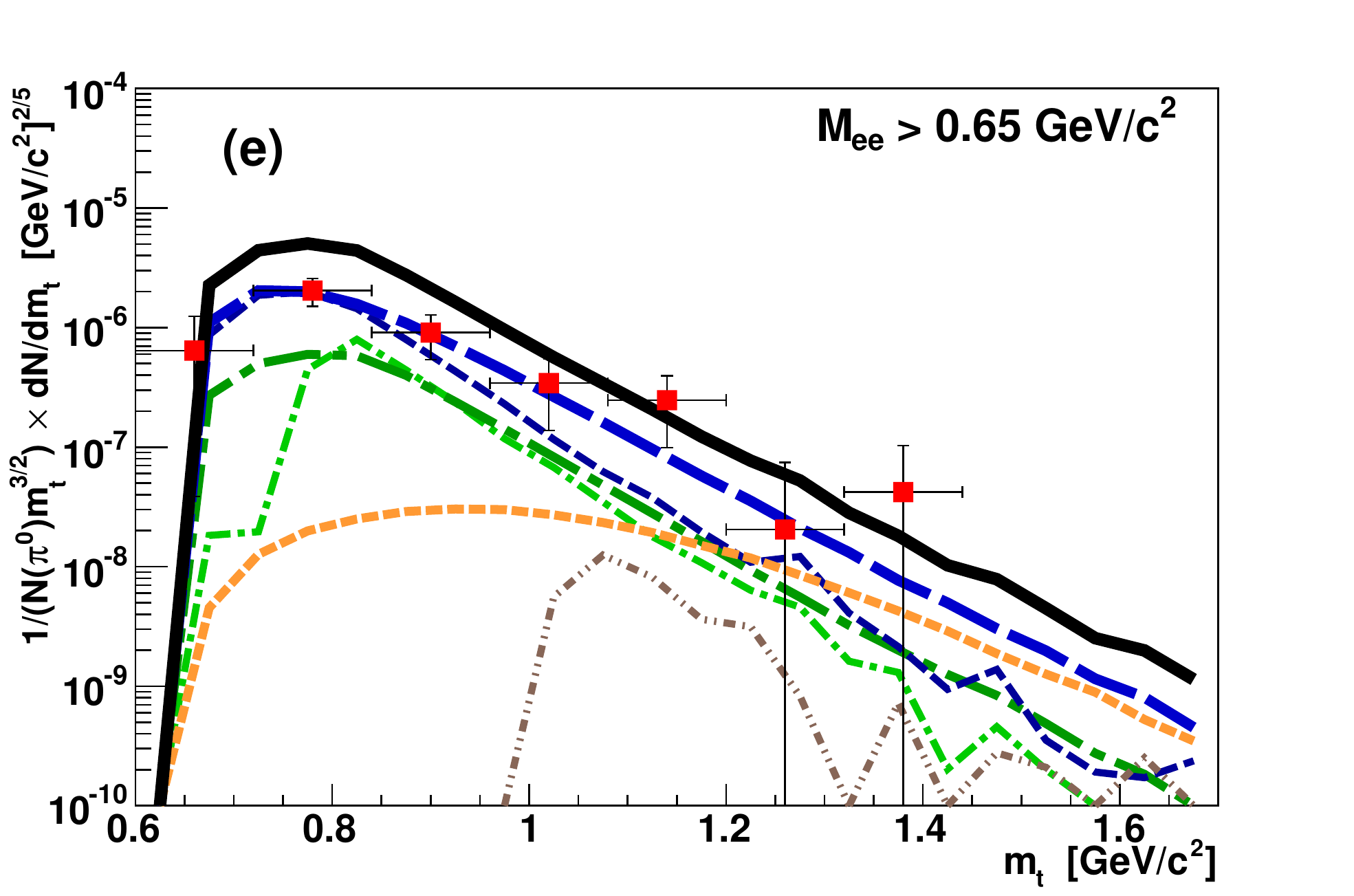}
\includegraphics[width=1.0\columnwidth]{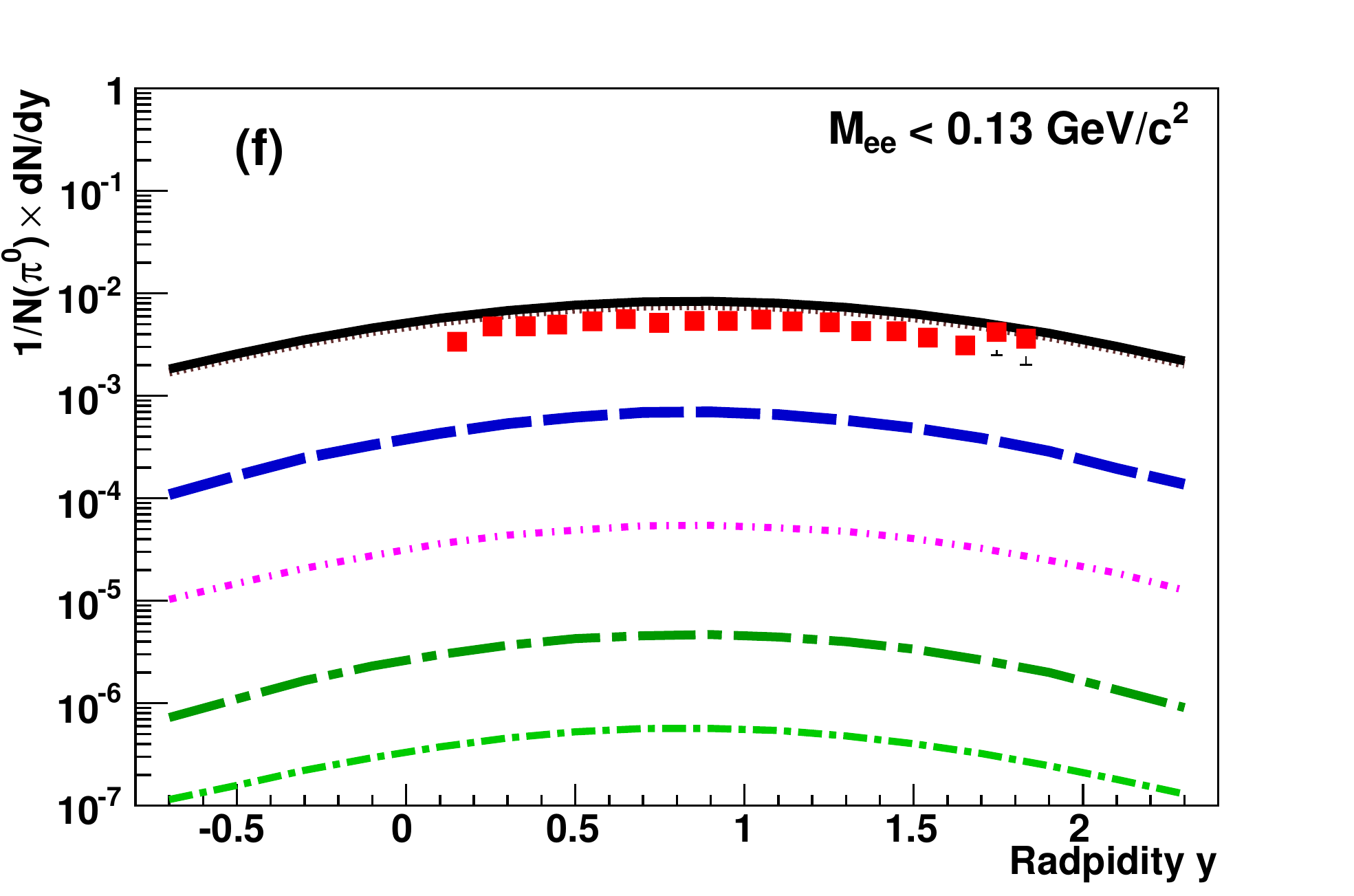}
\caption{(Color online) Transverse-mass spectra of the dielectron yield
  in different mass bins (a)-(e) and the dielectron rapidity spectrum of
  pairs with invariant mass $M_{ee} <$ 0.13 GeV/c$^{2}$ (f) for Ar+KCl
  collisions at $E_{\mathrm{lab}}=1.76 \,A$GeV. The results are compared to the
  experimental data from the HADES Collaboration
  \cite{Agakishiev:2011vf}. In contrast to Figures \ref{invmassCaAu} and
  \ref{vaccomp} the theoretical results here are not shown within the
  HADES acceptance, as these data are already fully corrected for
  acceptance and efficiency.}
\label{mtCa}
\end{figure*}
The resulting dielectron invariant-mass spectra for the two heavy-ion
reactions measured by the HADES Collaboration are presented in 
Fig.~\ref{invmass}. The left figure (a) shows the Ar+KCl results at 1.76\,$A$GeV
compared to the experimental data and the right figure (b) shows our prediction 
for the larger Au+Au system at 1.23\,$A$GeV (no dilepton data are 
published for the Au+Au measurement
yet). The coarse-graining results for the Ar+KCl case show good
agreement with the experimental data, especially the mass range between
0.2 and 0.6\,GeV/$c^{2}$ can be well described within this
approach. This is different from previous simulations which have 
indicated an excess of the
experimental outcome above the cocktail respectively the transport
calculations \cite{Agakishiev:2011vf,Weil:2012yg,Endres:2013nfa}. The dominant contribution stems from a broadened $\rho$. It is further noteworthy that
also the $\omega$ shows a non-negligible broadening in our model
results. A slight overestimation of the experimental dilepton yield shows up for
the low-mass region (below 150\,MeV/$c^{2}$) which is dominated by $\pi$
Dalitz decays, an effect that also has been found in other transport
calculations \cite{Weil:2012yg}. However, it remains unclear where this pion
excess stems from. As the dielectron yield is normalized to the total
$\pi^{0}$ number, theory and experiment should agree in the pion-dominated 
mass-region. Therefore the deviation may be due to a phase-space effect, 
connected to the geometrical detector acceptance. Another slight 
excess of the model results is manifest around the pole mass 
of the $\rho$. A significant part of the dileptons here 
stems from non-thermal $\rho$ mesons, which directly
come from the UrQMD calculations using the shining method. As for UrQMD the
$\rho$-production cross sections in the threshold region are known to be
slightly too large \cite{Endres:2013nfa}, this overestimation is probably
due to the non-thermal contribution. 

The medium effects become even stronger in the larger Au+Au system, as
can be seen in the right part (b) of Fig.~\ref{invmass}. The yield
from the thermal $\rho$ is higher at low masses compared to the Ar+KCl
reaction and also the thermal $\omega$ is slightly stronger here. This is in
line with the findings of the previous section \ref{ssec:ReacDyn}, where
it was pointed out that the hot and dense stage lives substancially longer 
in the Au+Au reaction. Consequently, the $\rho$ contribution is
larger for the Au+Au system as compared to the Ar+KCl reaction, 
especially at the low masses where the in-medium broadening comes 
into account. 

The dramatic effect of the presence of baryonic matter on the dilepton
spectra can be seen in Fig.~\ref{vaccomp}. Here the thermal
contributions to the $e^{+}e^{-}$ invariant mass spectrum of the $\rho$
and the $\omega$ for the case that no baryons and anti-baryons are present 
(i.e. the effective baryon density $\rho_{\mathrm{eff}}$ is put to zero) 
are compared to the results with the full medium effects. The baryon-driven effects are significant, with the broadening around the pole masses of the $\rho$ and 
$\omega$ meson and the strong increase in the low-mass dilepton yield, 
which is again more distinct for the larger Au+Au system as compared to Ar+KCl. Additionally the total sum for the two cases with and without baryonic effects is shown, including also all the UrQMD contributions (details are omitted for clarity). Here we find stronger differences between Ar+KCl and Au+Au, mainly due to the
relatively smaller contribution from the $\eta$ for Au+Au reactions,
compared to the stronger broadening of the $\rho$.

Looking not only at the invariant-mass spectra but also at the 
transverse-mass distributions in different mass bins in 
Fig.~\ref{mtCa} (a)-(e), one finds again a good agreement of our model 
calculations with the HADES data for Ar+KCl. While for the lowest mass bin
($M_{ee} < 0.13$\,GeV/c$^{2}$) the pion contribution dominates, the
thermal $\rho$ is the most significant contribution to the dileptons in
all other bins. For the highest mass bin ($M_{ee} > 0.65$\,GeV/c$^{2}$),
which includes the pole mass region of the $\rho$, one can observe a slight
overestimation of the total yield, similar to the one observed in the
invariant mass spectrum in this mass region (see Fig.~\ref{invmassCaAu}\,(a)).
Once again, we argue that this is due to the
high $\rho$-production cross section in the threshold region and
therefore stems from the non-thermal transport $\rho$. As one can see,
the thermal $\rho$ alone would describe the data very well.

Note that in some kinematic regions statistical fluctuations are seen in
the transverse mass spectra.  However, naturally this mainly affects
subleading contributions and the very high transverse masses. As the
production of certain particles is highly suppressed at SIS energies
(e.g., in case of the $\phi$ meson) it would need excessive computing
resources to remove these statistical fluctuations
completely. Nevertheless, the effect on the total yield is usually
rather small - especially compared to the uncertainty of the
experimental measurement. In general, an estimate of the global
(statistical and systematic) error of our dilepton calculations is
extremely difficult due to the many different parameters
(cross-sections, branching ratios, spectral functions, etc.)~which enter
the calculation and because for other sources (e.g., the filtering) the
error can hardly be quantified. At least for the long-lived
contributions (mainly $\pi$ and $\eta$) a comparison between different
transport models indicates that their contribution to the dilepton
spectra is quite well determined and does not differ much between the
models \cite{Bratkovskaya:2013vx,Schmidt:2008hm,
  Weil:2012ji,Weil:2014rea}. For the thermal contributions ($\rho$ and
$\omega$) the two main error sources are the uncertainties of the
spectral function and of the description of the reaction dynamics,
i.e., the time-evolution of temperature and chemical potential.

For the lowest mass bin, the HADES collaboration has also measured the
rapidity distribution of dielectrons. The results from the
coarse-grained simulations as well as the data points are shown in
Fig.~\ref{mtCa}\,(f). The shape of the spectrum is
reproduced well, but with some 20\% excess above the data points which
was already visible in the invariant mass spectrum. Note that the
rapidity values are for the laboratory frame, i.e. for
$E_{\mathrm{lab}}=1.76$\,$A$GeV mid-rapidity corresponds to a value of
$y_{0}=0.86$.

\subsubsection{DLS Results}
\begin{figure}
\includegraphics[width=1.0\columnwidth]{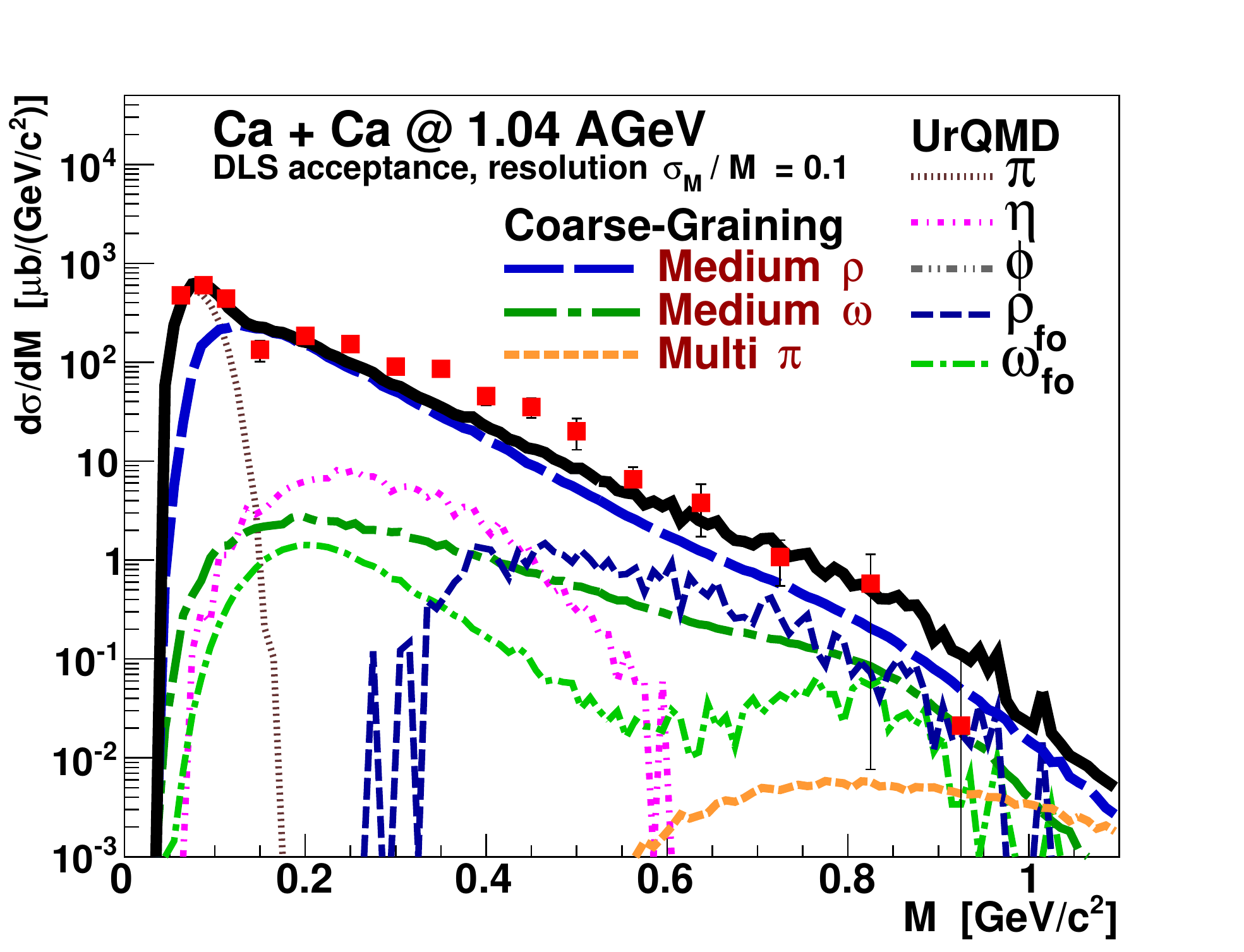}
\caption{(Color online) Invariant-mass spectrum of the dielectron yield
  for Ca+Ca collisions at $E_{\mathrm{lab}}=1.04 \,A$GeV within the experimental
  acceptance. The result is compared to the data from the DLS
  Collaboration \cite{Porter:1997rc}.}
\label{DLS}
\end{figure}
\begin{figure*}
\includegraphics[width=1.0\columnwidth]{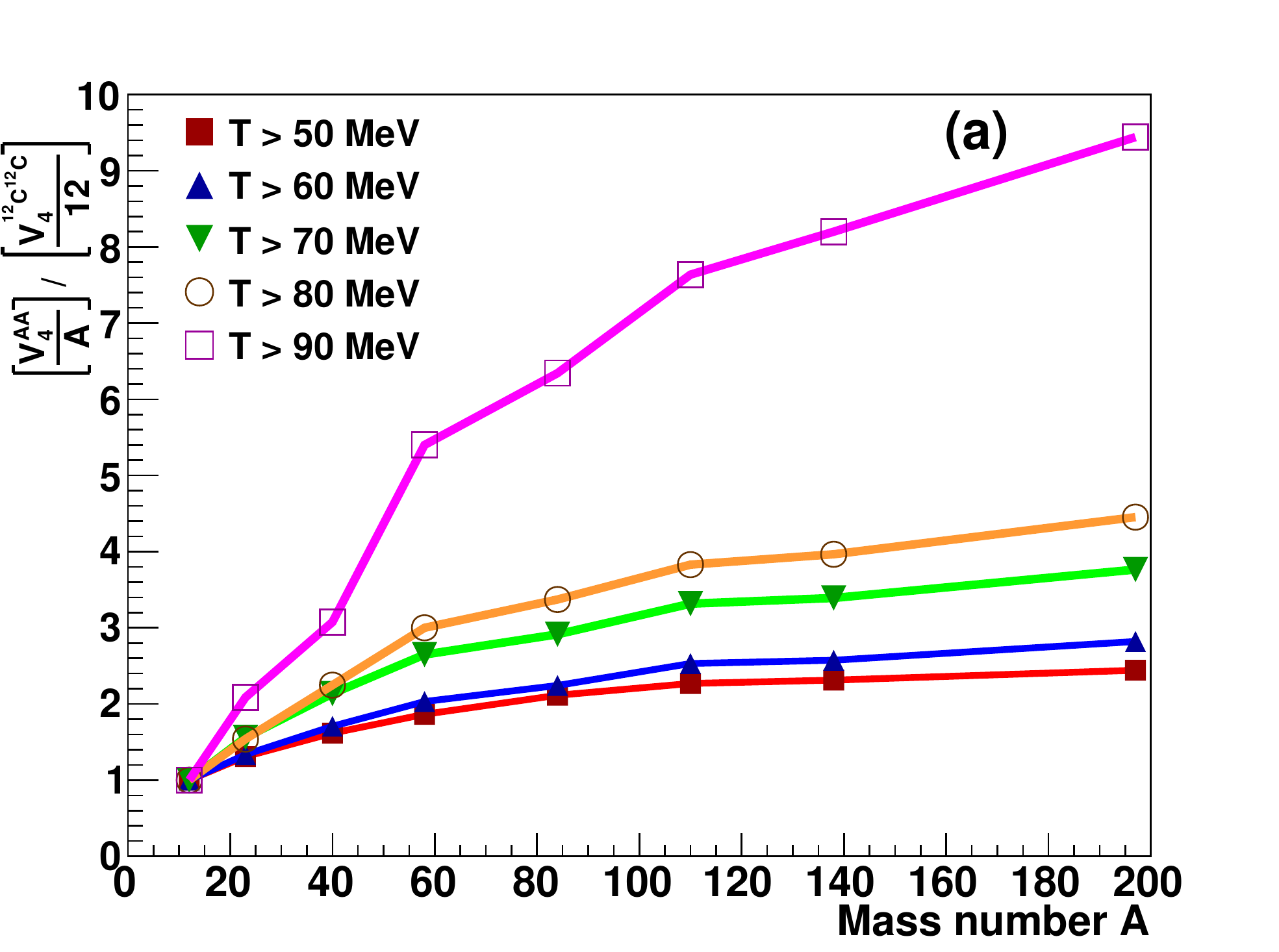}
\includegraphics[width=1.0\columnwidth]{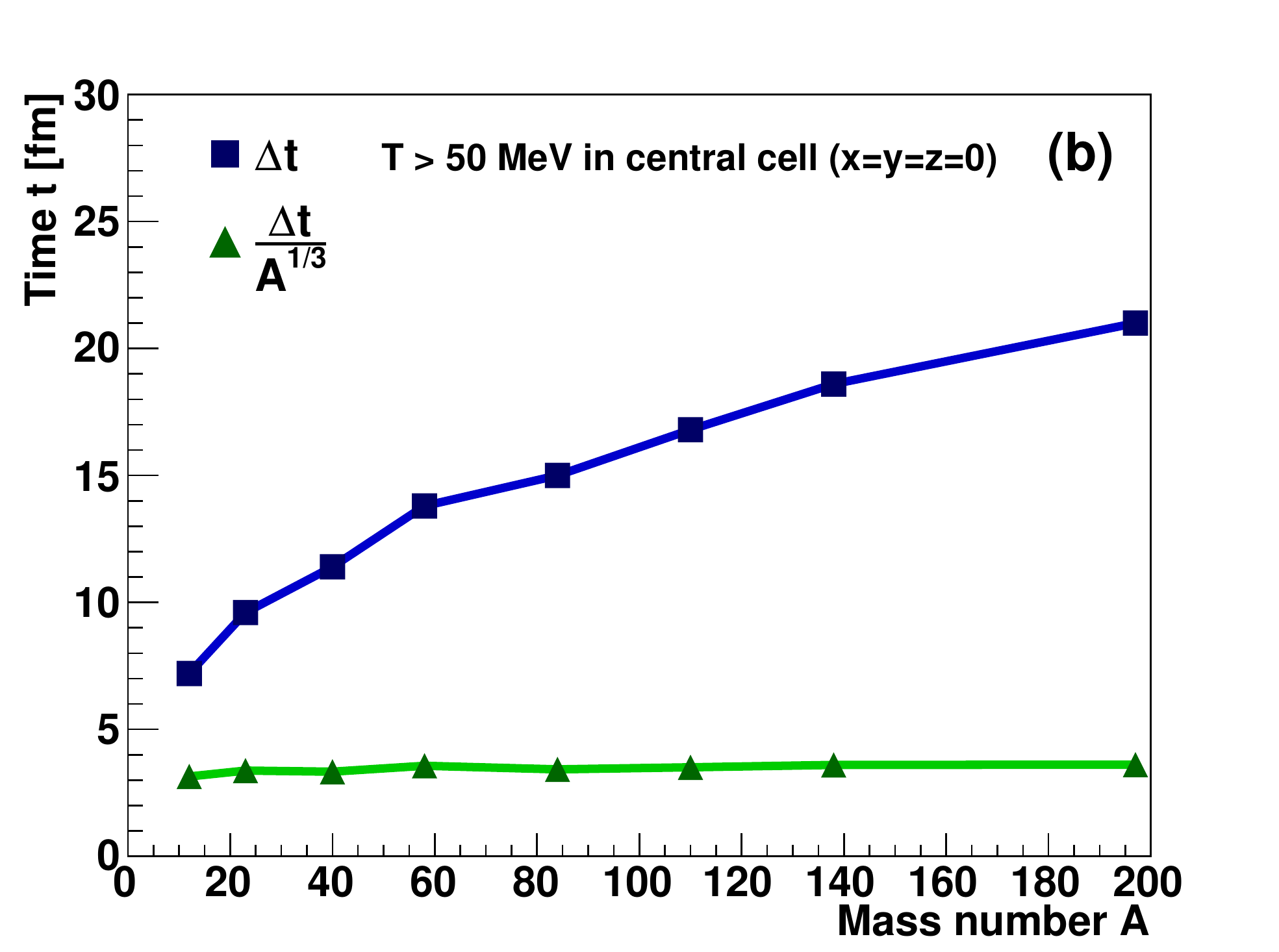}
\caption{(Color online) (a) Ratio of the thermal four-volume $V_{4}$ for
  different temperatures to the mass number $A$ of the colliding
  nuclei. The results are normalized to the ratio obtained with
  $^{12}\mathrm{C}+^{12}\mathrm{C}$ collisions. (b) Time duration over which the central
  cell of the coarse-graining grid (for $x=y=z=0$) emits thermal
  dileptons, i.e., for this period of time the central cell has a
  temperature which is above 50 MeV (blue squares) and the same scaled
  by $A^{1/3}$ (green triangles). Both plots (a) and (b) show the
  results in dependence on mass number $A$ for central collisions and a
  collision energy $E_{\mathrm{lab}} = 1.76$\,$A$GeV.}
\label{volratio}
\end{figure*}
Although the DLS measurement was done with a smaller acceptance and
lower resolution than in the more recent analyses by the HADES
Collaboration, it is nevertheless interesting to compare our results
also with these heavy-ion data. The system Ca+Ca which was measured by
DLS is comparable to the Ar+KCl reaction, as mentioned above. However,
the DLS Collaboration performed the measurement at a lower energy of
$E_{\mathrm{lab}}=1.04$\,$A$GeV. In  Fig.~\ref{DLS} the coarse-grained UrQMD
results for the according invariant-mass spectrum is shown within the
DLS acceptance, together with the experimental data points
\cite{Porter:1997rc}. In general the spectrum does not differ strongly
from the simulations for HADES, but the peaks -- especially in case of
the $\omega$ -- are more smeared out here, which is due to the low mass
resolution of only $\sigma_{M}/M=10\%$ of the detector, and the
acceptance at low masses is suppressed compared to the HADES
measurements. (For comparison, the mass resolution of the HADES
experiment is roughly 2\% in the region of the $\rho$ and $\omega$ pole
masses \cite{Galatyuk:2014oaa}.) While the overall description of the
experimental spectrum with the present model is relatively good, a
slight excess of the data above our model curve is present in the mass
range from 0.2 up to 0.6 GeV/c$^{2}$. This is in contrast to the
findings for the HADES case, where we could describe this mass region
quite accurately.

What might be the reasons for this deviation? It is possible that at
this very low energy other processes presently not considered in our
model might become more dominant here, e.g., an explicit bremsstrahlung
contribution (note, however, that some brems\-strah\-lung effects are
already considered within the in-medium spectral functions). As it has
been shown that the importance of brems\-strah\-lung contributions increases
with decreasing collision energy, these effects might be more significant 
than in Ar+KCl reactions at the slightly higher energies used for the HADES
experiment. Nevertheless, there are still uncertainties within the
different model calculations for the brems\-strah\-lung contribution,
differing by a factor 2--4 \cite{Kaptari:2005qz,Shyam:2010vr}. A main
problem here is the correct determination of the overall effect of the
different interfering processes, which is highly non-trivial. Another issue is that the lower
collision energy also results in slightly lower temperatures peaking 
around 80 MeV, which is indicating less thermalization of the
colliding system. Besides, one has to take into account that the DLS
experiment with its two-arm set-up had a more limited acceptance as
compared to the HADES detector. This makes it difficult to draw clear
quantitative conclusions from the comparison to the data. Additionally,
the lack of a measured impact-parameter distribution hampers precise
calculations within the coarse-graining approach, as the medium effects
can be quite sensitive to the centrality of the collision.

However, it was shown (at least for C+C reactions) that the HADES
results agree very well with those measured by the DLS collaboration if
the HADES data are filtered with the DLS acceptance
\cite{Agakishiev:2007ts}. So these data should provide a good additional
check for theoretical models in spite of their lower accuracy. Overall,
the agreement of our results with the data is quite good, especially considering that previous pure
transport calculations with the UrQMD model (without bremsstrahlung)
clearly underestimated the Ca+Ca data from DLS by a factor of 5-10 in
the mass range from 0.2 to 0.6\,GeV/c$^{2}$ \cite{Ernst:1997yy}.

\subsection{\label{syssize} System size and lifetime effects on thermal
dilepton emission} 

\begin{figure*}
\includegraphics[width=1.0\columnwidth]{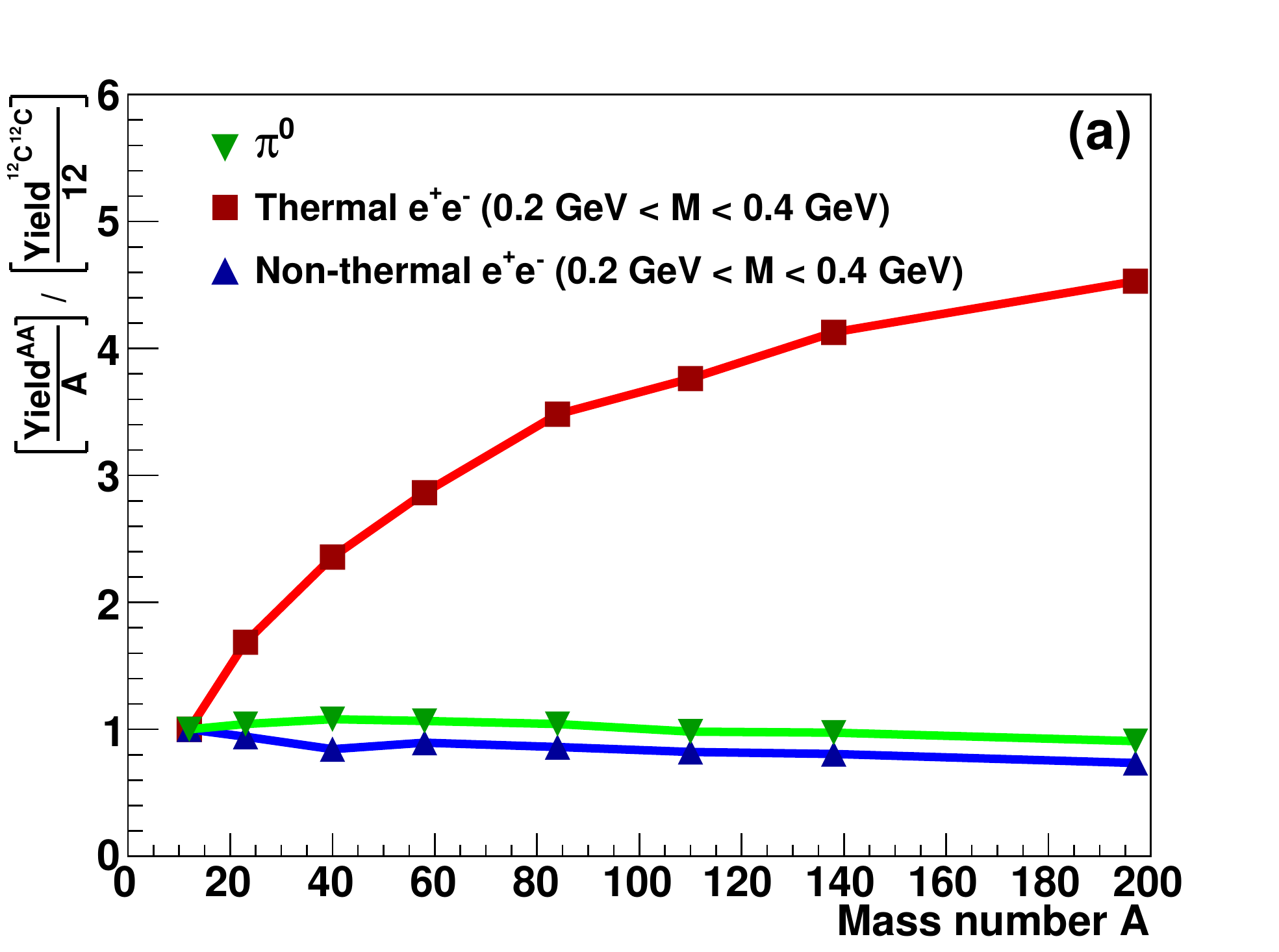}
\includegraphics[width=1.0\columnwidth]{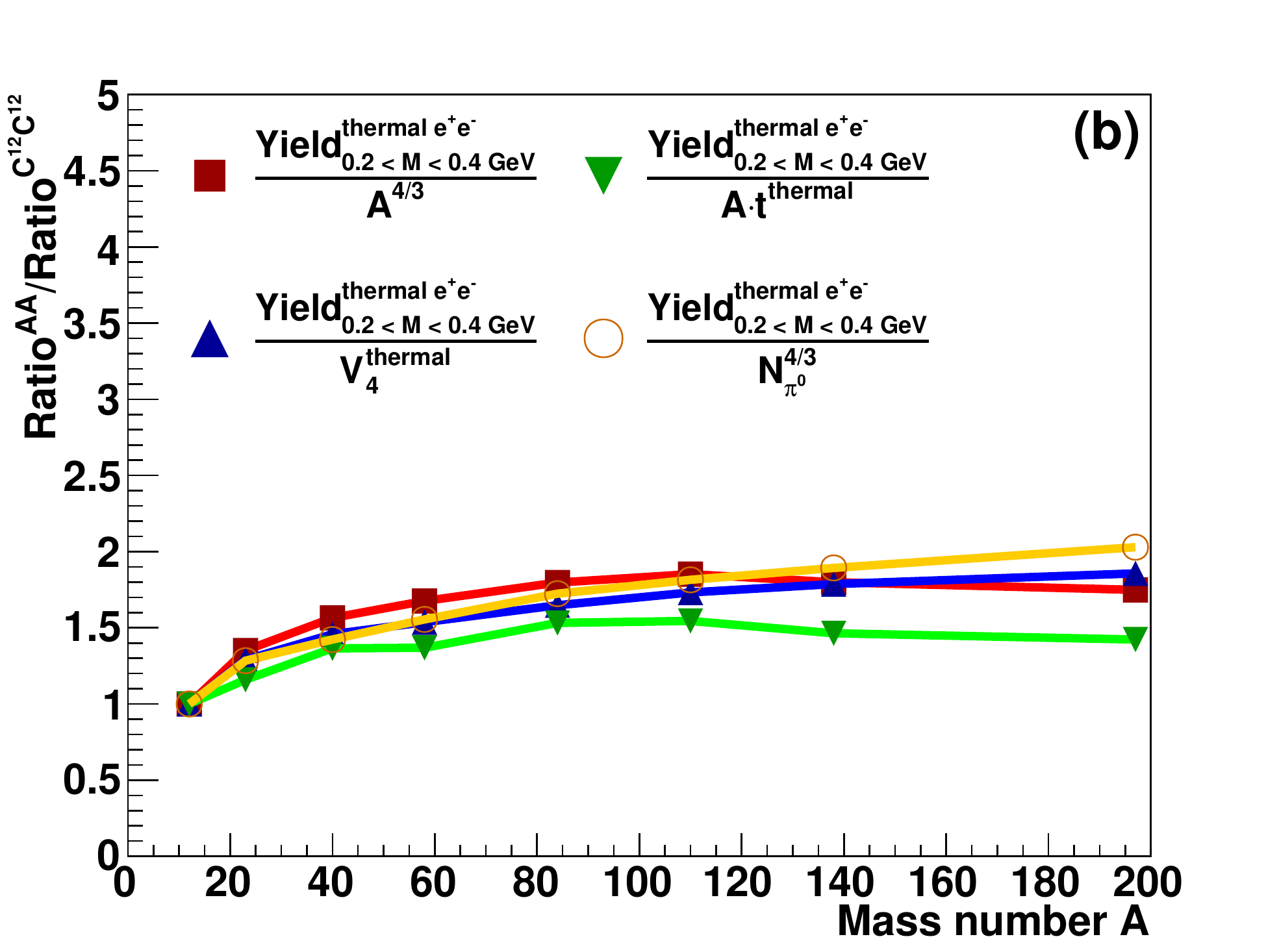}
\caption{(Color online) (a) Ratio of the thermal (red squares) and
  non-thermal dilepton yield (blue triangles) in the invariant mass
  range from 0.2 to 0.4\,GeV/$c^{2}$ to the mass number $A$ of the
  colliding nuclei; the same ratio is plotted for the total number of
  $\pi^{0}$ (green triangles). The results are normalized to the ratio
  obtained with $^{12}\mathrm{C}+^{12}\mathrm{C}$ collisions. (b) Ratio of the thermal dilepton yield in the invariant mass range from 0.2 to 0.4\,GeV/$c^{2}$ to $A^{4/3}$ (red squares), to the thermal four-volume (blue triangles), to the product of $A$ and the time
  $t^{\mathrm{thermal}}$ in which the central cell (for $x=y=z=0$) emits
  thermal dileptons, i.e, is at $T > 50$\,MeV (green triangles), and to the number of $\pi^{0}$ produced in the reaction scaled with an exponent 4/3 (orange circles). All ratios are normalized to the result obtained for
  $^{12}\mathrm{C}+^{12}\mathrm{C}$. Both plots (a) and (b) show the
  results for central collisions and a collision energy
  $E_{\mathrm{lab}} = 1.76$\,$A$GeV.}
\label{yieldratio}
\end{figure*}
\begin{figure}
\includegraphics[width=1.0\columnwidth]{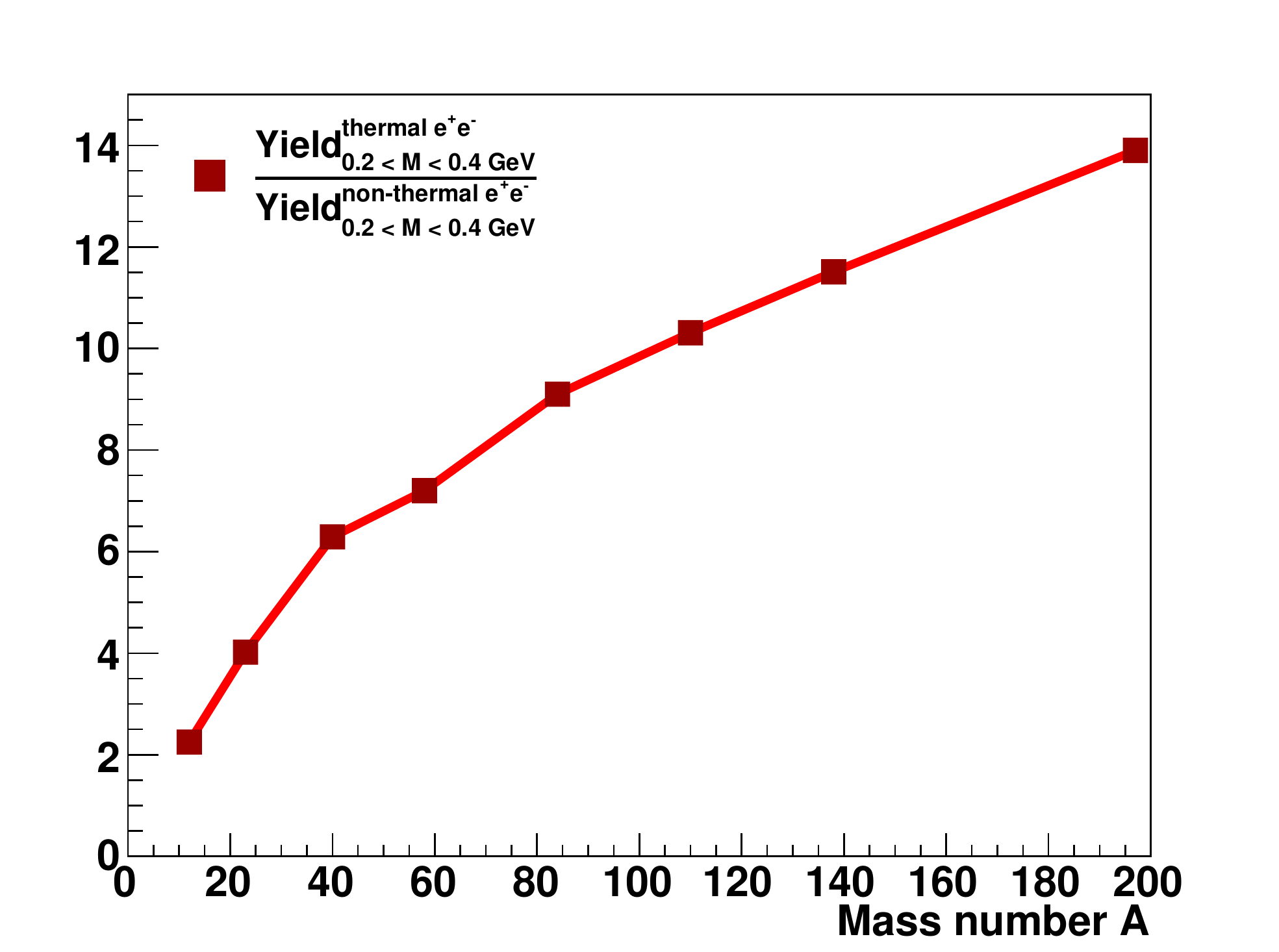}
\caption{(Color online) Ratio of the thermal to the non-thermal dilepton
  yield in the mass range $0.2\,\GeV/c^{2} < M_{ee} < 0.4 \,\GeV/c^{2}$
  for A+A collisions at $E_{\mathrm{lab}} = 1.76$\,$A$GeV. This ratio
  depicts the ``excess'' of the thermal dilepton yield with regard to
  the cocktail contributions. The results are shown in dependence on the
  mass number $A$ of the colliding nuclei.}
\label{yieldexcess}
\end{figure}
The results of the previous section indicate that the size of the
colliding nuclei (and in consequence also the size and duration of the
hot and dense fireball created thereby) largely influence the thermal
dilepton yield. This encourages a systematic study of the system size
dependence for thermal and non-thermal contributions and also raises the
question, up to which point the assumption of a thermalized system seems
reasonable. For this purpose we compare different systems from C+C to
Au+Au in central collisions at an energy of
$E_{\mathrm{lab}}=1.76$\,$A$GeV.

Fig.~\ref{volratio}\,(a) shows the ratio of the thermal four-volume
$V_{4}$ for different temperatures to the mass number $A$ of the
colliding nuclei. The results are normalized to the ratio obtained with
$^{12}\mathrm{C}+^{12}\mathrm{C}$ collisions.  It is noteworthy here
that the four-volume of the hottest cells ($T > 90$\,MeV) shows a much
stronger increase than the overall thermal volume (i.e., for
$T > 50$\,MeV). However, only the relative increase is shown in this
plot. In direct comparison the number of cells with highest temperatures
is very small compared to the total volume (approximately $1/30$ for
C+C). It is furthermore striking that the increase of the thermal
four-volume is not proportional to $A$, but it shows a stronger increase
for larger systems. This is not surprising, as $A$ is only a measure for
the volume of the colliding nuclei. In studies with the statistical
model it was shown that at SIS energies the thermal
freeze-out volume is closely related to the initial overlap volume of
the system created in $A-A$ collisions \cite{Cleymans:1998yb,
  BraunMunzinger:2003zd}. However, the thermal four-volume is also determined by the lifetime of the fireball. As it is difficult to determine some
kind of an average lifetime within the coarse-graining approach, we
concentrate on the central cell of the grid and assume that the time for
which it remains at an temperature greater than 50\,MeV should be to
first order a good approximation of the overall thermal lifetime. 
Fig.~\ref{volratio}\,(b) shows the mentioned time duration over which the
central cell emits thermal dileptons in dependence on $A$. Note that
this duration approximately scales with $A^{1/3}$, i.e.~with
the diameter of the nuclei. Obviously the main influence on the lifetime
at those low energies seems to be the time the two nuclei need to
traverse each other.

The question now is which influence the evolution of the four-volume has
on the production of lepton pairs. In Fig.~\ref{yieldratio}\,(a) the
ratio of thermal and non-thermal dilepton yield as well as the total
$\pi^{0}$ yield in the invariant mass range from 0.2 to 0.4\,GeV/$c^{2}$
in relation to the mass number $A$ of the colliding nuclei is shown. The
results are -- again -- normalized to the ratio obtained with C+C
collisions. While the ratio of the $\pi^{0}$ number as well as the
non-thermal dileptons to the mass number remains roughly 1 for all system
sizes (i.e. the non-thermal dilepton and $\pi^{0}$ yield increases
linearly with $A$), the thermal yield shows a significantly stronger
rise. This finding is in line with the larger thermal excess 
found in our calculations in Au+Au compared to Ar+KCl reactions.

If one now compares the thermal yield with the
corresponding total thermal four-volume (which is all cells with
$T > 50$\,MeV), as shown in Fig.~\ref{yieldratio}\,(b), the finding
indicates that the ratio between both shows only a very slight increase
and remains almost constant, independent of the system size. The same is
found if one calculates the relation between the thermal yield and the
product of $A$ with the thermal lifetime of the central cell (compare
Fig.\,\ref{volratio}\,(b)) or directly looks at the ratio between the
thermal dilepton yield and mass number $A$ respectively the number of
produced neutral pions $N_{\pi^{0}}$ scaled by an exponent
$\alpha=4/3$. All those quantities give an approximate measure of the
thermal four-volume at the low energies considered here. At higher
collision energies as, e.g., obtained at the CERN SPS or at RHIC (where
the whole fireball is more pion-dominated rather than baryon-dominated)
one would still expect an increase with $N_{\pi^{0}}^{\alpha}$, but not
any longer with $A^{\alpha}$. However, note that in these cases the
nuclei will traverse each other much faster and one will also find a
significant transverse expansion of the fireball, so that the diameter
of the nuclei can no longer be considered as a rough measure of the
lifetime; consequently the parameter $\alpha$ might be different
here. Nevertheless, in a different model calculation \cite{Rapp:2013nxa}
it was found that the thermal dilepton yield at top RHIC energy scales
with the number of charged particles as $N_{\mathrm{ch}}^{\alpha}$, where
$\alpha$ is found to take a value of approximately $1.45$. This result
is not so far from the value obtained within our simple picture -- at a
completely different energy regime.

In consequence, we learn from these results that the non-thermal
dilepton contributions (from long-lived 
me\-sons as the $\pi^{0}$ and the
$\eta$, respectively from the freeze-out $\rho$ and $\omega$) increase
with $A$ and therefore directly with the volume of the colliding
system. This is due to the fact that these contributions reflect the
final hadronic composition, after the whole reaction dynamics has
ended. Therefore the time-evolution of the system is irrelevant here. In
contrast, the thermal dilepton emission does not only increase with the
volume, but also with the time in which the colliding system remains in
a hot and dense stage. This is obvious, as the thermally emitted
dileptons will escape the fireball unscathed, while all hadronic
particles undergo processes as rescattering and reabsorption. In consequence, the thermal yield increases roughly proportional to $V\cdot \Delta t \sim A \cdot A^{1/3}=A^{4/3}$.
Exactly due to the different mechanisms contributing to thermal and non-thermal dilepton yields, one observes a significant increase of their ratio when going to larger system sizes. This ratio is shown (for the
invariant mass range form 0.2 to 0.4\,GeV/$c^{2}$) in Fig.~\ref{yieldexcess} 
with is value increasing from slightly over two
for C+C up to 14 for Au+Au.

The present results qualitatively agree with a recent study on the same
issue performed within different microscopic transport models, where the
total dilepton yield was found not to scale with $A$ or the number of
neutral pions, but showing a stronger increase due to the complicated
dynamics of the reaction \cite{Bratkovskaya:2013vx}. There, from a
microscopic point of view, it was also argued that the time evolution of
the reaction is the main reason for the increasing dilepton yield in
larger systems. If the dense phase with many binary scatterings lasts
longer, according to this picture, a larger brems\-strah\-lung contribution
(which is proportional to the number of collisions) and the repeated regeneration of $\Delta$ resonances raises the dilepton yield. However, the
contribution of other baryonic resonances was not explicitly considered there.
\section{\label{sec:3} Conclusions \& Outlook} 

We have presented results on dilepton production in 
\linebreak
Ar+KCl and Au+Au
collisions at GSI SIS\,18 and in 
\linebreak
Ca+Ca collisions at BEVALAC
energies. The results are obtained using a coarse-grained microscopic
transport approach and employing state-of-the-art spectral functions. With
this approach the experimental dilepton spectra in heavy-ion collisions
at $E_{\mathrm{lab}}=1$-2\,$A$GeV can be successfully described. The
model represents a third way to explore the dynamics of heavy-ion
reactions. In contrast to hydrodynamic/thermal fireball calculations or
microscopic transport simulations it allows for a consistent treatment
of both, high-energy and low-energy collisions.

Our results show that the dominant in-medium effect, which is also visible
in the dilepton spectra, stems from a strong broadening of the spectral
shape of the $\rho$ meson due to the high baryon density in the
fireball. This causes a melting of the $\rho$ peak and results in an
enormous increase of the dilepton production below the pole mass. The
reason for this is mainly the interaction with the baryonic resonances,
especially the $\Delta$ and $N^{*}_{1520}$, which are included in the
spectral function and give significant additional strength to the $\rho$
contribution at these low masses. The effect is much stronger at SIS
energies than for RHIC or SPS. Here, the baryon-chemical potential
remains very high for a significant time. This is visible
in the dilepton spectra which always represent a four-volume 
integral over the whole space-time evolution. In the
present study we also find a significant broadening of the $\omega$
meson at SIS energies. 

Furthermore, the influence of the size of the colliding system on the
thermal dilepton yield was studied. While we find the non-thermal
contribution in the mass range from 0.2 to 0.4\,GeV/$c^{2}$ (which stems
mainly from the long-lived $\eta$-meson but also from the ``freeze-out''
$\rho$ and $\omega$) to scale linearly with the mass number $A$, a
stronger increase of the thermal $\rho$ and $\omega$ yield is
observed. Their contribution to the dilepton yield scales with
$A t^{\mathrm{thermal}}$, i.e., the system's volume multiplied with the
lifetime of the thermal stage. Since $t^{\mathrm{thermal}}$ increases
with the diameter of the nuclei which is proportional to $A^{1/3}$, we
argue that the thermal dilepton emission should scale with $A^{4/3}$. As the number
of neutral pions directly scales with $A$, we find the thermal dilepton yield
also increasing with $N_{\pi^{0}}^{4/3}$. For future studies it might be 
quite instructive to check whether this proportionality still holds for higher
collision energies.

It is interesting to compare our findings with the results of pure
transport calculations. There, in recent calculations the dilepton
excess in the invariant-mass spectra above the cocktail were mainly
explained by two different effects: (i) Bremsstrahlung contributions 
and (ii) Dalitz decays of baryonic resonances
\cite{Bratkovskaya:2013vx,Weil:2014rea}. It is important to understand
that both effects are also included in the spectral functions applied
here. These processes, as implemented in the transport models, correspond
to cuts of the in-medium self-energy diagrams of the $\rho$ meson. For
example, a contribution to the self-energy from a $\Delta$-hole
excitation would be represented in the transport model by the process
$\Delta\rightarrow \rho N \rightarrow \gamma^{*}N$, assuming strict
vector meson dominance. However, due to the completely different
character of the approaches it is difficult to compare the single
contributions quantitatively. Also note that the self-energies in
hadronic many-body quantum-field theory approach correspond to a
\emph{coherent} superposition of the various scattering processes, while
in transport approaches naturally the various processes are summed
\emph{incoherently}. A future detailed analysis of the self-energy
contributions and their relative strengths might be fruitful to better
understand how far the both approaches, i.e., the transport (kinetic)
and the equilibrium thermal quantum-field theory description of the
dilepton-emission rates at such low energies agree. One should keep in
mind that the same microscopic scattering and decay processes are
underlying both approaches.

Apart from this, the main outcome of the present investigation is the
possibility of a consistent description of dilepton production from SPS
down to SIS energies within the same model. In both energy regimes the
spectra can be described reasonably well by the assumption of medium
modifications of the vector mesons' spectral properties, whereby the
$\rho$ plays the most significant role.

In the future, the CBM experiment at FAIR will offer the possibility to
study medium effects in a collision-energy range, where this kind of
study has not been carried out yet. With high (net-)baryon densities but
also temperatures reaching up to or above the critical temperature
$T_{c}$, this will be a further test for the spectral functions and
theoretical models. Besides, the high luminosities expected at FAIR
might enable to perform more detailed and systematic studies, as, e.g.,
the effect of various different system sizes on the dilepton yield or a
study for different centralities, to obtain more information on the
evolution of the reaction dynamics.
\\
\section*{Acknowledgements}
The authors thank the HADES Collaboration and especially Tetyana
Galatyuk for providing the acceptance filters and experimental
data. Furthermore, we acknowledge Ralf Rapp for providing the
parametrization of the spectral functions and many fruitful
discussions. This work was supported by the Bundesministerium f{\"u}r 
Bildung und Forschung, Germany (BMBF), the Hessian Initiative for 
Excellence (LOEWE) through the Helmholtz International Center for
FAIR (HIC for FAIR) and the Helmholtz Association through the
Helmholtz Research School for Quark-Matter Studies (H-QM).

\bibliography{Bibliothek}

\end{document}